\documentclass{aipproc} 
\layoutstyle{6s}
\usepackage{amsmath,amssymb,graphicx}
\graphicspath{{Eps/}}

\input{alphabet}
\def\bdoc{\begin{document}}             \def\edoc{\end{document}}
\def\babs{\begin{abstract}}             \def\eabs{\end{abstract}}
\def\bcc{\begin{center}}                \def\ecc{\end{center}}
\def\ben{\begin{enumerate}}             \def\een{\end{enumerate}}
\def\bit{\begin{itemize}}               \def\eit{\end{itemize}}
\def\bpic{\begin{picture}}              \def\epic{\end{picture}}
\def\beq{\begin{equation}}              \def\eeq{\end{equation}}
\def\barr{\begin{array}}                \def\earr{\end{array}}
\def\btab{\begin{tabular}}              \def\etab{\end{tabular}}
\def\btabu{\begin{tabular}}             \def\etabu{\end{tabular}}
\def\bcc{\begin{center}}                \def\ecc{\end{center}}
\def\beqn{\begin{eqnarray}}             \def\eeqn{\end{eqnarray}}
\def\beqnx{\begin{eqnarray*}}           \def\eeqnx{\end{eqnarray*}}
\def\bfig{\begin{figure}}               \def\efig{\end{figure}}
\def\bver{\begin{verb}}						 \def\ever{\end{verb}}

\def\d#1{\,\mbox{d} #1}
\def\disp#1{\displaystyle{#1}}

\def\blue#1{{\color{blue}#1}}
\def\red#1{{\color{red}#1}}
\def\green#1{{\color{green}#1}}
\def\magenta#1{{\color{magenta}#1}}
\def\cyan#1{{\color{cyan}#1}}
\def\titre#1{\setlength{\shadowsize}{4pt}\centerline{\shadowbox{\blue{\mbox{~~\sc #1~~}}}}}

\def\iii{\int_{-\infty}^{+\infty}}
\def\izi{\int_{0}^{\infty}}
\def\izpi{\int_{0}^{\pi}}
\def\izdpi{\int_{0}^{2\pi}}
\def\intd{\int\kern-.8em\int}
\def\intt{\int\kern-.8em\int\kern-.8em\int}
\def\intg{\int\kern-1.1em\int}

\def\xh{\widehat{x}}
\def\thetah{\widehat{\theta}}
\def\betah{\widehat{\beta}}

\def\xbh{\widehat{\xb}}
\def\thetabh{\widehat{\thetab}}
\def\betabh{\widehat{\betab}}

\def\xbhk{\widehat{\xb}^{k}}
\def\thetahk{\widehat{\theta}^{k}}
\def\betahk{\widehat{\beta}^{k}}

\def\xbhkp{\widehat{\xb}^{k+1}}
\def\thetahkp{\widehat{\theta}^{k+1}}
\def\betahkp{\widehat{\beta}^{k+1}}

\def\thetabhk{\widehat{\thetab}^{k}}
\def\betabhk{\widehat{\betab}^{k}}

\def\thetabhkp{\widehat{\thetab}^{k+1}}
\def\betabhkp{\widehat{\betab}^{k+1}}

\def\thetamin{\theta_{\mbox{\tiny min}}}
\def\thetamax{\theta_{\mbox{\tiny max}}}
\def\betamin{\beta_{\mbox{\tiny min}}}
\def\betamax{\beta_{\mbox{\tiny max}}}

\def\ra{\rightarrow}
\def\la{\leftarrow}
\def\da{\downarrow}
\def\ua{\uparrow}

\def\Ra{\Rightarrow}
\def\La{\Leftarrow}
\def\Da{\Downarrow}
\def\Ua{\Uparrow}

\def\lra{\longrightarrow}
\def\lla{\longleftarrow}
\def\Lra{\Longrightarrow}
\def\Lla{\longleftarrow}

\def\lrarr{\leftrightarrow}
\def\Lrarr{\Leftrightarrow}
\def\udarr{\updownarrow}
\def\Uparr{\Updownarrow}

\def\expf#1{\exp\left[ {#1} \right]}

\def\argmax#1#2{\arg\max_{#1}\left\{ #2 \right\}}
\def\argmin#1#2{\arg\min_{#1}\left\{ #2 \right\}}

\def\Prob#1{\mbox{Pr}\left\{#1\right\}}
\def\var#1{\mbox{Var}\left\{#1\right\}}
\def\cov#1{\mbox{Cov}\left\{#1\right\}}
\def\corr#1{\mbox{Corr}\left\{#1\right\}}
\def\trace#1{\mbox{Tr}\left\{#1\right\}}
\def\rang#1{\mbox{rang}\left\{#1\right\}}
\def\det#1{\mbox{d\'et}\left\{#1\right\}}

\def\gbu{\underline{\gb}}
\def\fbu{\underline{\fb}}
\def\zbu{\underline{\zb}}
\def\epsilonbu{\underline{\epsilonb}}
\def\thetabu{\underline{\thetab}}

\def\sigmae{{\sigma_{\epsilon}}}
\def\Sigmae{{\Sigmab_{\epsilonb}}}
\def\Sigmaf{{\Sigmab_{\fb}}}
\def\Sigmag{{\Sigmab_{\gb}}}

\def\fbuh{\widehat{\fbu}}
\def\zbuh{\widehat{\fbu}}
\def\thetabuh{\widehat{\thetabu}}

\def\fbh{\widehat{\fb}}
\def\Pbh{\widehat{\Pb}}
\def\Sigmabh{\widehat{\Sigmab}}
\def\Abh{\widehat{\Ab}}
\def\Bbh{\widehat{\Bb}}
\def\thetabh{\widehat{\thetab}}
\def\Rbe{\Rb_{\epsilon}}
\def\Rbeh{\widehat{\Rbe}}

\def\Rjk{{R_j}_k}
\def\fbik{{\fb_i}_k}
\def\fbjk{{\fb_j}_k}

\def\mik{{m_i}_k}
\def\sigmaik{{\sigma_i^2}_k}
\def\Sigmaik{{\Sigmab_i}_k}
\def\alphaik{{\alpha_i}_k}
\def\betaik{{\beta_i}_k}

\def\muik{{\mu_i}_k}
\def\vik{{v_i}_k}

\def\mjk{{m_j}_k}
\def\sigmajk{{\sigma_j^2}_k}
\def\Sigmajk{{\Sigmab_j}_k}
\def\alphajk{{\alpha_j}_k}
\def\betajk{{\beta_j}_k}

\def\mujk{{\mu_j}_k}
\def\vjk{{v_j}_k}
\def\njk{{n_j}_k}

\def\alphae{\alpha^{\epsilon}}
\def\betae{\beta^{\epsilon}}
\def\alphaez{\alpha^{\epsilon}_{0}}
\def\betaez{\beta^{\epsilon}_{0}}
\def\alphaei{\alpha^{\epsilon}_{i}}
\def\betaei{\beta^{\epsilon}_{i}}
\def\alphaeiz{\alpha^{\epsilon}_{i0}}
\def\betaeiz{\beta^{\epsilon}_{i0}}

\def\mbz{{{\mb}_z}}
\def\Sigmabz{{{\Sigmab}_z}}
\def\diag#1{\mbox{diag}\left[#1\right]}

\def\sigmah{\widehat{\sigma}}
\def\fh{\widehat{f}}
\def\sigmaz{\sigma_z}

\def\blue#1{{\color{blue}#1}}
\def\red#1{{\color{red}#1}}
\def\green#1{{\color{green}#1}}

\def\sumi{\sum_{i=1}^{M} }
\def\sumj{\sum_{j=1}^{N} }
\def\lambdah{\wh{\lambda}}
\def\lambdabh{\wh{\lambdab}}

\def\det#1{\hbox{det}\left(#1\right)}
\def\defined{\,\shortstack{$\triangle$\\ =}\,}

\def\zmat{\left[\matrix{0}\right]}
\def\det#1{\hbox{det}(#1)}
\def\th{\widehat{t}}
\def\xh{\widehat{x}}
\def\lambdah{\widehat{\lambda}}
\def\muh{\widehat{\mu}}
\def\sigmah{\widehat{\sigma}}
\def\rhoh{\widehat{\rho}}
\def\betah{\widehat{\beta}}
\def\alphah{\widehat{\alpha}}
\def\tbh{\widehat{\tb}}
\def\mubh{\widehat{\mub}}
\def\sigmabh{\widehat{\sigmab}}
\def\rhobh{\widehat{\rhob}}
\def\oneb{\mbox{\bf 1}}

\def\bm#1{\mbox{\boldmath $#1$}}
\def\zerob{{\bf 0}}
\def\oneb{{\bf 1}}
\def\intg{\int\kern-1.1em\int}
\def\expf#1{\exp\left[ {#1} \right]}

\def\gbu{\underline{\gb}}
\def\fbu{\underline{\fb}}
\def\zbu{\underline{\zb}}
\def\epsilonbu{\underline{\epsilonb}}
\def\uthetab{\underline{\thetab}}

\def\sigmae{{\sigma_{\epsilon}}}
\def\Sigmae{{\Sigma_{\epsilonb}}}
\def\Sigmabe{{\Sigmab_{\epsilonb}}}
\def\Sigmaf{{\Sigmab_{\fb}}}
\def\Sigmag{{\Sigmab_{\gb}}}

\def\fbuh{\widehat{\fbu}}
\def\zbuh{\widehat{\fbu}}
\def\uthetabh{\widehat{\uthetab}}

\def\fbh{\widehat{\fb}}
\def\Sigmabh{\widehat{\Sigmab}}
\def\Abh{\widehat{\Ab}}
\def\thetabh{\widehat{\thetab}}

\def\Rjk{{R_j}_k}
\def\fbik{{\fb_i}_k}
\def\fbjk{{\fb_j}_k}

\def\mik{{m_i}_k}
\def\sigmaik{{\sigma_i^2}_k}
\def\Sigmaik{{\Sigmab_i}_k}
\def\alphaik{{\alpha_i}_k}
\def\betaik{{\beta_i}_k}

\def\muik{{\mu_i}_k}
\def\vik{{v_i}_k}

\def\mjk{{m_j}_k}
\def\sigmajk{{\sigma_j^2}_k}
\def\Sigmajk{{\Sigmab_j}_k}
\def\alphajk{{\alpha_j}_k}
\def\betajk{{\beta_j}_k}

\def\mujk{{\mu_j}_k}
\def\vjk{{v_j}_k}
\def\njk{{n_j}_k}

\def\alphae{\alpha^{\epsilon}}
\def\betae{\beta^{\epsilon}}
\def\alphaez{\alpha^{\epsilon}_{0}}
\def\betaez{\beta^{\epsilon}_{0}}
\def\alphaei{\alpha^{\epsilon}_{i}}
\def\betaei{\beta^{\epsilon}_{i}}
\def\alphaeiz{\alpha^{\epsilon}_{i0}}
\def\betaeiz{\beta^{\epsilon}_{i0}}
\def\alphaeiz{\alpha^{\epsilon}_{i0}}
\def\betaeiz{\beta^{\epsilon}_{i0}}
\def\alphaeiz{\alpha^{\epsilon}_{i0}}
\def\betaeiz{\beta^{\epsilon}_{i0}}

\def\mbz{{{\mb}_z}}
\def\Sigmabz{{{\Sigmab}_z}}

\def\vect{\mbox{vect}}
\def\lra{\longrightarrow}

\def\gir{g_i(\rb)}
\def\fjr{f_j(\rb)}
\def\zjr{z_j(\rb)}
\def\epsilonir{\epsilon_i(\rb)}
\def\gbr{\gb(\rb)}
\def\fbr{\fb(\rb)}
\def\zbr{\zb(\rb)}
\def\epsilonbr{\epsilonb(\rb)}
\def\fbhr{\fbh(\rb)}
\def\Sigmabhr{\Sigmabh(\rb)}

\def\mbzr{\mb_{z(\rb)}}
\def\Sigmabzr{\Sigmab_{z(\rb)}}
\def\Sigmagbzr{{\Sigmab_g}_{z(\rb)}}

\def\mjzjr#1{{m_{#1}}_{z_{#1}(\rb)}}
\def\sigmajzjr#1{{\sigma_{#1}}_{z_{#1}(\rb)}}

\def\Beta{B}
\def\Betab{\bm{\Beta}}
\def\Betabe{\bm{\Beta}_{\epsilonb}}

\def\fh{\widehat{f}}
\def\sigmah{\widehat{\sigma}}
\def\sigmaei{\sigma_{\epsilon_i}^2}

\def\thetah{\widehat{\theta}}
\def\sigmah{\widehat{\sigma}}

\def\Ftxm{\widetilde{F}_{X}(x_-)}
\def\Ftxp{\widetilde{F}_{X}(x_+)}

\def\NU{{\cal V}}

\def\tFx{\widetilde{F}_{X}}

\def\Fx{F_{X}(x)}
\def\fx{f_{X}(x)}

\def\Fxv{F_{X}(x)}
\def\fxv{f_{X}(x)}

\def\Fxi{{F}_{X_i}(x_i)}
\def\fxi{{f}_{X_i}(x_i)}

\def\Gnu{G_{\NU}(\nu)}
\def\gnu{g_{\NU}(\nu)}

\def\Fnu{F_{\NU}(\nu)}
\def\fnu{f_{\NU}(\nu)}

\def\Finnu#1{F^{-1}_{\NU}(#1)}

\def\Gnua#1{G_{\NU}(#1)}
\def\Gnub#1{G_{\NU}^{-1}(#1)}
\def\Gnuh{G_{\NU}^{-1}(1/2)}

\def\Ftx{\widetilde{F}_{X}(x)}
\def\ftx{\widetilde{f}_{X}(x)}

\def\Ftxi{\widetilde{F}_{X_i}(x_i)}
\def\ftxi{\widetilde{f}_{X_i}(x_i)}

\def\Ftxv{\widetilde{F}_{X}(x)}
\def\FtX#1{\widetilde{F}_{X}(#1)}
\def\ftxv{\widetilde{f}_{X}(x)}

\def\fxnu{f_{X|\NU}(x|\nu)}
\def\fxNU{f_{X|\NU}(x|\NU)}
\def\FxNU{F_{X|\NU}(x|\NU)}

\def\fxNUtheta{f_{X|\NU,\theta}(x|\NU,\theta)}
\def\FxNUtheta{F_{X|\NU,\theta}(x|\NU,\theta)}

\def\Fxnu{F_{X|\NU}(x|\nu)}
\def\Fxnui#1{F_{X|\NU}(x|\nu_#1)}

\def\FxN#1{F_{X|\NU}(#1)}
\def\FxNU{F_{X|\NU}(x|\NU)}
\def\Fxnun#1{F_{X|\NU}(x|#1)}
\def\Fxinun#1{F_{X_i|\NU}(x_i|#1)}
\def\fxnu{f_{X|\NU}(x|\nu)}
\def\fxNU{f_{X|\NU}(x|\NU)}

\def\Fxvnu{F_{X|\NU}(x|\nu)}
\def\Fxvnun#1{F_{X|\NU}(x|#1)}
\def\FxvNU{F_{X|\NU}(x|\NU)}

\def\FXvnu{F_{X|\NU}}

\def\fxvnu{f_{X|\NU}(x|\nu)}
\def\fxvNU{f_{X|\NU}(x|\NU)}

\def\FXvNU#1{{F}_{X|\NU}(#1|\NU)}
\def\FXvn#1{{F}_{X|\NU}(x|#1)}

\def\Fxtheta{F_{X|\theta}(x|\theta)}
\def\fxtheta{f_{X|\theta}(x|\theta)}

\def\Fxvtheta{F_{X|\theta}(x|\theta)}
\def\fxvtheta{f_{X|\theta}(x|\theta)}

\def\Ftxtheta{\widetilde{F}_{X|\theta}(x|\theta)}
\def\ftxtheta{\widetilde{f}_{X|\theta}(x|\theta)}

\def\Ftxvtheta{\widetilde{F}_{X|\theta}(x|\theta)}
\def\ftxvtheta{\widetilde{f}_{X|\theta}(x|\theta)}

\def\Fxnutheta{F_{X|\NU,\theta}(x|\nu,\theta)}
\def\fxnutheta{f_{X|\NU,\theta}(x|\nu,\theta)}

\def\ftheta{f_{\Theta}(\theta)}
\def\fthetaxnuz{f_{\Theta|X,\nu_0}(\theta|x,\nu_0)}
\def\fthetax{f_{\Theta|X}(\theta|x)}

\def\Fxvnutheta{F_{X|\NU,\theta}(x|\nu,\theta)}
\def\FxvNUtheta{F_{X|\NU,\theta}(x|\NU,\theta)}
\def\fxvnutheta{f_{X|\NU,\theta}(x|\nu,\theta)}
\def\Ftxvnutheta{\widetilde{F}_{X|\theta}(x|\theta)}
\def\ftxvnutheta{\widetilde{f}_{X|\theta}(x|\theta)}

\def\FxvNUm{F_{X|\NU}(x_-|\NU)}
\def\FxvNUp{F_{X|\NU}(x_+|\NU)}

\def\Fxinu{F_{X_i|\NU}(x_i|\nu)}
\def\fxinu{f_{X_i|\NU}(x_i|\nu)}

\def\Ftxi{\widetilde{F}_{X_i}(x_i)}
\def\ftxi{\widetilde{f}_{X_i}(x_i)}

\def\fxitheta{f_{X_i|\theta}(x_i|\theta)}
\def\ftxitheta{\widetilde{f}_{X_i|\theta}(x_i|\theta)}

\def\thetah{\widehat{\theta}}
\def\thetat{\widetilde{\theta}}

\def\lnuxv{l_{\nu|X}(\nu|x)}

\def\Lnuxva#1{L_{\nu|X}(#1 | x)}
\def\lnuxva#1{l_{\nu|X}(#1 | x)}

\def\Prob#1{\mbox{P}\left(#1\right)}

\def\lthetaxvnuz{l(\theta)}

\def\lthetaxvnuz{l_{\theta|x,\nu_0}(\theta | x,\nu_0)}
\def\fxnuztheta{f_{X|\nu_0,\theta}(x | \nu_0,\theta)}
\def\fxinuztheta{f_{X|\nu_0,\theta}(x_i | \nu_0,\theta)}
\def\fxvtheta{f_{X|\theta}(x | \theta)}
\def\lthetaxv{l_{\theta|X}(\theta | x)}
\def\ltthetaxv{\widetilde{l}_{\theta|X}(\theta | x)}

\def\ltheta{l(\theta )}
\def\lttheta{\widetilde{l}(\theta)}
\def\Lnu{L(\nu )}
\def\Lnui#1{L(\nu_#1)}
\def\Linu{L_i(\nu )}
\def\Linu{L_i(\nu )}
\def\LNU{L(\NU )}
\def\Lin#1{L^{-1}(#1)}
\def\L#1{L(#1)}
\def\Li#1{L_i(#1)}
\def\FNU#1{F_\NU(#1)}
\def\FinNU#1{F^{-1}_\NU(#1)}

\def\ftthetax{\tilde{f}_{\Theta|X}(\theta|x)}
\def\ER{R}

\def\Ndd#1#2#3{{\cal N}\left( #1 ;~ #2, #3 \right)}
\def\Cdd#1#2#3{{\cal C}\left( #1 ;~ #2, #3 \right)}
\def\Sdd#1#2#3#4{{\cal S}\left( #1 ;~ #2, #3, #4 \right)}
\def\Edd#1#2{{\cal E}\left( #1 ;~ #2 \right)}
\def\DEdd#1#2{{\cal DE}\left( #1 ;~ #2 \right)}
\def\Gdd#1#2#3{{\cal G}\left( #1 ;~ #2, #3 \right)}
\def\IGdd#1#2#3{{\cal IG}\left( #1 ;~ #2, #3 \right)}

\def\Xwr{X(\omega,\rb)}
\def\xwr{x(\omega,\rb)}

\def\muzw{\mu_{z}(\omega)}
\def\muzwr{\mu_{z(\rb)}(\omega)}
\def\szw{\sigma_{z}^2(\omega)}
\def\szwr{\sigma_{z(\rb)}^2(\omega)}
\def\pzw{p_{z}(\omega)}
\def\pzwr{p_{z(\rb)}(\omega)}

\def\hszw{\widehat{\sigma_{z}^2}(\omega)}
\def\hmuzw{\widehat{\mu}_{z}(\omega)}
\def\hpzw{\widehat{p}_{z}(\omega)}

\def\muzwrl{\mu_{z(\rb)}(\omega-l)}

\def\pxwrcz{p\left(\xwr ~|~ z\right)}
\def\pxwr{p\left(\xwr\right)}

\def\Xbw{\Xb(\omega)}
\def\xbw{\xb(\omega)}
\def\uXb{\underline{\Xb}}
\def\uxb{\underline{\xb}}
\def\pxbw{p(\xbw)}
\def\puxb{p(\uxb)}
\def\pzb{P(\zb)}

\def\Szw{S_z(\omega)}
\def\mubw{\mub(\omega)}

\def\Srw{S_{\rb}(\omega)}
\def\Swr{S_{\omega}(\rb)}
\def\Sbw{\Sb(\omega)}
\def\sbw{\sb(\omega)}
\def\Sbr{\Sb(\rb)}
\def\mubw{\mub(\omega)}
\def\cov#1{\mbox{cov}[#1]}

\def\pdf{\emph{pdf}}
\def\pmf{\emph{pmf}}
\def\dpdx#1#2{\frac{\partial #1}{\partial #2}}

\def\REM#1{}

\def\XXwr{X(\omega,\rb)}
\def\xxwr{x(\omega,\rb)}

\def\Xwr{X_{\omega}(\rb)}
\def\xwr{x_{\omega}(\rb)}
\def\Xrw{X_{\rb}(\omega)}
\def\xrw{x_{\rb}(\omega)}

\def\Ywr{Y_{\omega}(\rb)}
\def\ywr{y_{\omega}(\rb)}
\def\Yrw{Y_{\rb}(\omega)}
\def\yrw{y_{\rb}(\omega)}

\def\Ewr{E_{\omega}(\rb)}
\def\ewr{\epsilon_{\omega}(\rb)}
\def\Erw{E_{\rb}(\omega)}
\def\erw{\epsilon_{\rb}(\omega)}

\def\Xbr{\Xb(\rb)}
\def\xbr{\xb(\rb)}
\def\Xbw{\Xb(\omega)}
\def\xbw{\xb(\omega)}

\def\uXb{\underline{\Xb}}
\def\uxb{\underline{\xb}}
\def\uXbz{\underline{\Xb}_z}
\def\uxbz{\underline{\xb}_z}

\def\uthetab{\underline{\thetab}}
\def\uthetabz{\underline{\thetab}_z}

\def\Ywr{Y_{\omega}(\rb)}
\def\ywr{y_{\omega}(\rb)}
\def\Yrw{Y_{\rb}(\omega)}
\def\yrw{y_{\rb}(\omega)}
\def\Ybr{\Yb(\rb)}
\def\ybr{\yb(\rb)}
\def\Ybw{\Yb(\omega)}
\def\ybw{\yb(\omega)}
\def\uYb{\underline{\Yb}}
\def\uyb{\underline{\yb}}
\def\uYbz{\underline{\Yb}_z}
\def\uybz{\underline{\yb}_z}

\def\Ewr{E_{\omega}(\rb)}
\def\ewr{\epsilon_{\omega}(\rb)}
\def\Erw{E_{\rb}(\omega)}
\def\erw{\epsilon_{\rb}(\omega)}
\def\Ebr{Eb(\rb)}
\def\ebr{\epsilonb(\rb)}
\def\Ebw{Eb(\omega)}
\def\ebw{\epsilonb(\omega)}
\def\uEb{\underline{\Eb}}

\def\Erwp{E_{\rb}(\omega')}
\def\Xrwp{X_{\rb}(\omega')}

\def\xwmr{x_{\omega-1}(\rb)}
\def\muzwm{\mu_{z}(\omega-1)}

\def\rhoz{\rho_z}
\def\sez{\sigma_{\eta_z}^2}

\def\xwrp{x_{\omega}(\rb')}
\def\xrwp{x_{\rb}(\omega')}
\def\xwpr{x_{\omega'}(\rb)}
\def\xwprp{x_{\omega'}(\rb')}
\def\Xwrp{x_{\omega}(\rb')}
\def\Xwrp{X_{\omega}(\rb')}
\def\Xwpr{X_{\omega'}(\rb)}
\def\Xwprp{X_{\omega'}(\rb')}

\def\sigmaed{\sigmae^2}
\def\sigmaew{\sigmae(\omega)}
\def\sigmaedw{\sigmaed(\omega)}

\def\Sigmabe{\Sigmab_{\epsilon}}
\def\sigmabh{\widehat{\Sigmab}}
\def\xbh{\widehat{\xb}}
\def\zbh{\widehat{\zb}}
\def\qbh{\widehat{\qb}}
\def\sbh{\widehat{\sb}}

\def\argmax#1#2{\arg\max_{#1}\left\{#2\right\}}
\def\argmin#1#2{\arg\min_{#1}\left\{#2\right\}}
\def\espx#1#2{\mbox{E}_{#1}\left\{#2\right\}}
\def\esp#1{\mbox{E}\left\{#1\right\}}
\def\d#1{\mbox{~d}#1}

\def\Tr#1{\mbox{Tr}\left[#1\right]}

\def\syslins#1#2#3#4{
\bpic(150,40)
  \put(5,15){\makebox(15,0){#1}}
  \put(20,15){\vector(1,0){10}}
  \put(30,7){\framebox(30,15){#2}}
  \put(60,15){\vector(1,0){10}}
  \put(75,15){\circle{10}}
  \put(75,15){\makebox(0,0){+}}
  \put(80,15){\vector(1,0){10}}
  \put(125,15){\makebox(15,0){#4}}
  \put(75,30){\vector(0,-1){10}}
  \put(68,32){\makebox(30,0){#3}}
\epic
}
\def\syslinsa#1#2#3#4{
\bpic(90,40)
  \put(5,15){\makebox(10,0){#1}}
  \put(15,15){\vector(1,0){10}}
  \put(25,7){\framebox(20,15){#2}}
  \put(45,15){\vector(1,0){10}}
  \put(60,15){\circle{10}}
  \put(60,15){\makebox(0,0){+}}
  \put(65,15){\vector(1,0){10}}
  \put(75,15){\makebox(10,0){#4}}
  \put(60,30){\vector(0,-1){10}}
  \put(60,32){\makebox(10,0){#3}}
\epic
}
\def\inversions#1#2#3{
\begin{picture}(100,40)
  \put(0,15){\makebox(15,0){#1}}
  \put(20,15){\vector(1,0){10}}
  \put(30,7){\framebox(40,15){#2}}
  \put(70,15){\vector(1,0){10}}
  \put(85,15){\makebox(15,0){#3}}
\end{picture}
}

\def\tomoXaz{
\begin{picture}(70,68)
  \put(15,35){\makebox(20,15){$\red{f(x,y)}$}}
  \put(-1,0){\axexy}
  \put(10,10){\objet}
  \put(5,5){\vector(1,1){55}}
  \put(55,55){\makebox(5,10){$r$}}
  \put(35,28){\makebox(5,10){$\phi$}}
  \put(60,8){\makebox(10,15){$\bullet$D}}
  \put(60,0){\makebox(15,10){$\blue{g(r,\phi)}$}}
  \put(10,55){\makebox(10,10){S$\bullet$}}
  \put(62,15){\line(-1,1){45}}
  \put(40,-5){\proj}
\thinlines  \blue{\multiput(10,14)(5,0){9}{\line(0,1){40}}}
\thinlines  \blue{\multiput(8,14)(0,5){9}{\line(1,0){40}}}
\end{picture}
}

\def\tomoXay{
\begin{picture}(70,70)
  \put(15,35){\makebox(20,15){$f(x,y)$}}
  \put(-1,0){\axexy}
  \put(10,10){\objet}
  \put(5,5){\vector(1,1){55}}
  \put(55,55){\makebox(5,10){$r$}}
  \put(35,28){\makebox(5,10){$\phi$}}
  \put(60,8){\makebox(10,15){$\bullet$D}}
  \put(60,0){\makebox(15,10){$g(r,\phi)$}}
  \put(10,55){\makebox(10,10){S$\bullet$}}
  \put(62,15){\line(-1,1){45}}
  \put(40,-5){\proj}
\end{picture}
}

\def\slice{
\begin{picture}(140,70)
  \put(13,35){\makebox(20,15){\red{$f(x,y)$}}}
  \put(35,28){\makebox(5,10){$\phi$}}
  \put(80,10){\makebox(5,10){\blue{$g(r,\phi)$--FT--$G(\Omega,\phi)$}}}
  \put(0,0){\axexy}
  \put(0,0){\axers}
  \put(10,10){\objet}
  \put(35,28){\makebox(5,10){$\phi$}}
  \put(35,-5){\proj}
  \put(80,0){\axewxwy}
  \put(80,0){\axeaw}
  \put(90,30){\makebox(20,15){\red{$F(\omega_{x}, \omega_{y})$}}}
  \put(115,28){\makebox(5,10){$\phi$}}
  \put(96,15){\line(1,1){30}}
\end{picture}
}

\def\cresta{\tt    \setlength{\unitlength}{0.8pt}
\begin{picture}(120,120)
\thicklines   \multiput(0,27)(0,20){5}{\line(1,0){80}}
              \multiput(0,27)(20,0){5}{\line(0,1){80}}
              \put(61,33){\red{$f_N$}}
              \put(0,92){\red{$f_1$}}
              \put(40,72){\red{$f_j$}}
              \put(88,50){\blue{$g_{i}$}}
              \put(38,120){\blue{$H_{ij}$}}
              \put(-5,115){\line(3,-2){90}}
              \put(42,85){\blue{\line(3,-2){20}}}
\end{picture}}

\def\crestb{\tt    \setlength{\unitlength}{0.8pt}
\begin{picture}(120,120)
\thicklines   \multiput(0,27)(0,20){5}{\line(1,0){80}}
              \multiput(0,27)(20,0){5}{\line(0,1){80}}
              \put(61,33){\red{$f_N$}}
              \put(0,92){\red{$f_1$}}
              \put(40,72){\red{$f_j$}}
              \put(0,12){\blue{$g_1$}}
              \put(60,12){\blue{$g_m$}}
              \put(88,33){\blue{$g_{m+1}$}}
              \put(88,75){\blue{$g_i$}}
              \put(88,95){\blue{$g_M$}}
              \put(-5,75){\blue{\line(1,0){90}}}
\end{picture}}

\def\homedir{/home/djafari/}
\def\Picfs{\homedir Tex/Ali/Picfs/}
\def\Epsfs{\homedir Tex/Ali/Eps/}
\def\sfEps{\homedir Matlab/sf2d/Eps/}
\def\decEps{\homedir Matlab/dec1d/Eps/}
\def\imEps{\homedir Matlab/dec2d/Eps/}
\def\matlabdat{\homedir Matlab/dat/}
\def\radonEps{\homedir Matlab/Radon/Eps/}
\def\Physics{\homedir Tex/Ali/Articles/Physics/}
\def\Maxent{\homedir Tex/Ali/Congres/Maxent96/}
\def\RappEps{\homedir Tex/Ali/Rapports/Rapp95/Eps/}
\def\ProbaEps{\homedir Tex/Ali/Proba/Eps/}

\def\Nc{{\cal N}}
\def\Vc{{\cal V}}
\def\thetab{\bold{\theta}}

\def\tzb{\widetilde{\zb}}
\def\tthetab{\widetilde{\thetab}}
\def\qh{\widehat{q}}

\title{Dirichlet or Potts ?}

\author{Ali Mohammad-Djafari}
 {
  address = {Laboratoire des Signaux et Syst\`emes,\linebreak 
  Unit\'e mixte de recherche 8506 (CNRS-Sup\'elec-UPS 11) \linebreak  
  Sup\'elec, Plateau de Moulon, 3 rue Juliot-Curie, 91192 Gif-sur-Yvette, France},
  email = {djafari@lss.supelec.fr}
 }
\date{}

\begin{abstract}
\noindent When modeling the distribution of a set of data by a mixture of Gaussians, there are two possibilities: 
~i) the classical one is using a set of parameters which are the proportions, the means and the variances; 
ii) the second is to consider the proportions as the probabilities of a discrete valued hidden variable. 
In the first case a usual prior distribution for the proportions is the Dirichlet which accounts for the fact that they have to sum up to one. 
In the second case, to each data is associated a hidden variable for which we consider two possibilities: 
a) assuming those variables to be i.i.d. We show then that this scheme is equivalent to the classical mixture model with Dirichlet prior; b) assuming a Markovian structure. Then we choose the simplest markovian model which is the Potts distribution. As we will see this model is more appropriate for the case where the data represents the pixels of an image for which the hidden variables represent a segmentation of that image. 
The main object of this paper is to give some details on these models and different algorithms used for their simulation and the estimation of their parameters. 
\end{abstract}

\begin{document}
\maketitle

\section{Introduction}
When modeling the distribution of a set of data $\xb=\{x_i, i=1,\cdots,N\}$ by a mixture of Gaussians (MoG), there are two possibilities: 
\bit
\item[ i)] The classical one is using a set of parameters which are the proportions   $\alphab=\{\alpha_k, k=1,\cdots,K\}$, the means $\mub=\{\mu_k, k=1,\cdots,K\}$ and the variances $\vb=\{v_k, k=1,\cdots,K\}$: 
\beq \label{model1}
p(x)=\sum_{k=1}^K \alpha_k \Nc(x|\mu_k, v_k)
\eeq
and the objective is the estimation of $K$ and the parameters 
$\thetab=\{\alphab,\mub, \vb\}$. 

\item[ii)] The second is to consider the proportions $\alpha_k$ as the probabilities of a discrete value hidden variable $Z$ whith $\alpha_k=P(Z=k)$: 
\beq \label{model2}
p(x)=\sum_{k=1}^K P(Z=k) \Nc(x|\mu_k, v_k)
\eeq
which implies that $p(x|Z=k)=\Nc(\mu_k, v_k)$.  
\eit

In the first case a usual prior distribution for $\alphab=\{\alpha_k, k=1,\cdots,K\}$ is the Dirichlet
\beq \label{Dirichlet}
\Dc(\alphab|\lambdab)=\frac{\Gamma\left(\sum_k \lambda_k\right)}{\prod_k \Gamma(\lambda_k)}\prod_{k=1}^K \alpha_k^{\lambda_k-1}
\eeq
which accounts  for the fact that $\sum_k \alpha_k=1$. 

In the second case, to each data $x_i$ is associated a discrete value hidden variable $Z_i$. The value $z_i\in{1,\cdots,K}$ which takes $Z_i$ is then the class label of the datum $x_i$. When $\xb=\{x_i, i=1,\cdots,N\}$ represent the pixels of an image, $\zb=\{z_i, i=1,\cdots,N\}$ represents its segmentation. 
Then, naturally, we consider two possibilities for the distribution of $\zb$: a) assuming the variables $Z_i$ to be i.i.d.; b) assuming that there is a spatial structure through the image pixel index $i$ and thus assigning them a Potts Markov distribution. 

In this paper we give some details on these models and different algorithms used 
for their simulation and the estimation of their parameters. 

\section{Maximum Likelihood and Bayesian approaches}
In model (\ref{model1}), the classical maximum likelihood (ML) method assumes that the data 
$\xb=\{x_i, i=1,\cdots,n\}$  
are i.i.d samples from (\ref{model1}) and thus 
\beq \label{likelihood}
\Lc(\xb|\thetab,K)=\prod_i p(x_i)=\prod_{i=1}^n \sum_{k=1}^K \alpha_k \Nc(x_i|\mu_k, v_k).
\eeq
Then, for a given $K$, the objective is the estimation of 
$\thetab=\{(\alpha_k,\mu_k,v_k), k=1,\cdots,K\}$ which is defined as 
\beq \label{ML}
\thetabh=\argmax{\thetab}{\Lc(\xb|\thetab)}
\eeq
It is important to note that, the likelihood expression can become degenerate in the sense that it may become unbounded for particular set of parameters and data \cite{Snoussi01c}. 
This makes the estimation of the parameters by this approach difficult. This is the reason for many authors to propose the penalized likelihood criteria to overcome this difficulty. The penalization term has the role to eliminate this degeneracy \cite{Champagnat95,Ridolfi00,Ciuperca03}.

In the Bayesian approach, one assigns priors $\pi(\thetab)$, finds the expression of the posterior 
\beq \label{Bayes}
p(\thetab|\xb)\propto \Lc(\xb|\thetab) \; \pi(\thetab)
\eeq
and then, an estimate $\thetabh$ is defined either as the MAP estimate:
\beq \label{MAP}
\thetabh=\argmax{\thetab}{\Lc(\xb|\thetab)\; \pi(\thetab)}
\eeq
or the posterior mean 
\beq \label{PM}
\thetabh=\int \thetab \; p(\thetab|\xb) \d{\thetab} 
=\frac{\int \thetab \; \Lc(\xb|\thetab)\; \pi(\thetab) \d{\thetab}}{ 
\int \Lc(\xb|\thetab)\; \pi(\thetab) \d{\thetab}} 
\eeq

The choice of the prior $\pi(\thetab)$ in the Bayesian approach for the MoG model has been 
the subject of interest for many Bayesian authors through the entropic or conjugate priors. 
Both approaches result to the same prior, at least for the proportion parameters $\alpha_k$ 
which is the Dirichlet prior~(\ref{Dirichlet}). 
The conjugate priors for the means are the Gaussians 
\beq 
\pi(\mu_k)=\Nc(\mu_k|\mu_0,v_0)
\eeq
and for the variances are the Inverse Gamma (IG). 
\beq 
\pi(v_k)=\Ic\Gc(v_k|\alpha_0,\beta_0)
\eeq

What is also interesting to note is that using the IG prior for the variances in the MAP estimate 
results exactly to the necessary penalization term in the ML approach which is needed to eliminate the degeneracy of the likelihood. 

Computing the ML solution (\ref{ML}) or the MAP solution (\ref{MAP}) can be done either 
directly or through an EM algorithm, but the PM solution (\ref{PM}) can not be obtained analytically and needs Mont\'e Carlo (MC) algorithms. It is curious to note that, in the EM algorithm as well as in the MC sampling methods, one introduces the notion of hidden variables which is the subject of the second case modeling. 

\section{Separable (Dirichlet) and Markovian (Potts) models for the hidden variables}
In model (\ref{model2}), to each data $x_i$ is associated a hidden variable $Z_i$ and the assumption is that the data $x_i$ is a sample from 
$p(x_i|Z_i=k)=\Nc(x_i|\mu_k, v_k), \forall i$ 
where the $Z_i$ can only take the values 
$k=1,\cdots,K$. 

Then if we assume $Z_i$ to be independent and identically distributed (iid):
\beq
P(Z_i=k)=\alpha_k, \forall i \quad \mbox{and}\quad P(Z_i=k,Z_j=l)=\alpha_k\alpha_l, \forall i,j 
\eeq
we can write
\beq \label{Dirichlet1}
P(\Zb=\zb|\alphab,K)=\prod_i \alpha_{z_i}=\prod_k \alpha_k^{\sum_i \delta(z_i-k)},
\eeq
which means that $\Zb$ is separable in $Z_i$, 
then we can find a link between the two models (\ref{model1}) and (\ref{model2}) which 
become equivalent with 
$\alpha_k=\frac{1}{n} \sum_{i=1}^n \delta(z_i-k)$. 

But if we assume that there are some structure (dependancy) in the hidden variables, 
then, we have to model them.  
The simplest model for such a structure is the Potts model: 

\beq \label{Potts1}
P(Z_i=z_i|Z_j=z_j, j\not= i)\propto \exp\left\{\gamma \sum_{j\in\Vc(i)} \delta(z_i-z_j)\right\}
\eeq
where $\Vc(i)$ represents the neighboring elements of $i$, for example 
$\Vc(i)=i-1$ or 
$\Vc(i)=\{i-1, i+1\}$ or in cases where $i$ represents the index of a pixel in an image, then 
$\Vc(i)$ represents the four nearest neigbors of that pixel. $\gamma$ is the Potts parameter. 

Using the equivalence of Gibbs and Markovian distributions, we can also write
\beq \label{Potts2}
\left\{\barr{lcl}
\pi(z_i|z_j,j\in\Vc(i),\gamma,K)&\propto& \exp\left\{\gamma \sum_{j\in\Vc(i)} \delta(z_i-z_j)\right\}\\
\pi(\zb|\gamma,K)&\propto& \exp\left\{\gamma \sum_i\sum_{j\in\Vc(i)} \delta(z_i-z_j)\right\}
\earr\right.
\eeq
where $\pi(z_i)$ stands in short for $P(Z_i=z_i)$ and $\pi(\zb)$ stands in short for $P(\Zb=\zb)$.

\section{Data classification and Image segmentation}
These two models have been used in many data classification or image segmentation where 
the $x_i$ represents either the grey level or the color components of the pixel $i$ and 
$z_i$ its class labels. The main objective of an image segmentation algorithm is the 
estimation of $z_i$. When the hyperparameters 
$K$, $\thetab=(\alpha_k,\mu_k,v_k), k=1,\cdots,K$ and $\gamma$ 
are not known and have also to be estimated, we say that we are in \emph{totally unsupervised} mode, when are known we are in \emph{totally supervised} mode and we say that we are in  
\emph{partially supervised} mode when some of those hyperparameters are fixed. A classical case 
is the one with fixed $K$. 

Assuming first $K$ known, we can write the following: 
\beq
p(\xb|\zb,\thetab,K)=\prod_{i\in\Rc} p(x_i|z_i)
=\prod_{i\in\Rc} \Nc(x_i|\mu_{z_i},\sigma_{z_i}^2)
=\prod_k \prod_{i\in\Rc_k} \Nc(x_i|\mu_{z_i},\sigma_{z_i}^2)
\eeq
where $\Rc=\{1,\cdots,n\}$ represents the set of all samples (all pixels positions of an image) and $\Rc_k=\{i : z_i=k\}$ represents the set of all samples who have the same label value $z_i=k$. Evidently, we assume that $\cup_k \Rc_k = \Rc$ which means that all samples are classified. 

\beq
p(\zb,\thetab|\xb,K,\gamma)\propto p(\xb|\zb,\thetab,K) \; \pi(\zb|\gamma,K) \; \pi(\thetab) 
\eeq
Then, one can try to estimate both $\zb$ and $\thetab$ from this expression either by alternate maximization:
\beq \label{AM1}
\left\{\barr{lcl}
\zbh&=&\argmax{\zb}{p(\zb,\thetabh|\xb,K,\gamma)}\\ 
\thetabh&=&\argmax{\thetab}{p(\zbh,\thetab|\xb,K,\gamma)}
\earr\right.
\eeq
or by first estimating $\theta$ and then using it for the estimation of $\zb$:
\beq \label{GML}
\thetabh=\argmax{\thetab}{p(\thetab|\xb,K,\gamma)} \lra 
\zbh=\argmax{\zb}{p(\zb|\thetabh,\xb,K,\gamma)}.
\eeq
However, the first step of this second approach cannot be done explicitly and needs an iterative 
algorithm using the hidden variables $\zb$ as the missing data. The Bayesian EM algorithm has particularly been developped for this:
\beq \label{BEM}
\left\{\barr{llcl}
\mbox{E step:} & Q(\thetab|\thetabh^{(t)})&=&\esp{\ln p(\Zb,\thetab|\xb,K,\gamma) | \thetabh^{(t)}}\\ 
\mbox{M step:} & \thetabh^{(t+1)}&=&\argmax{\thetab}{Q(\thetab|\thetabh^{(t)})}
\earr\right.
\eeq

The full Bayesian approach consists in exploring the whole posterior probability distribution by generating samples from it. This can be done through a Gibbs sampling algorithm:
\beq \label{GS1}
\left\{\barr{lcl}
\zbh&\sim&{p(\zb|\thetabh,\xb,K,\gamma)}\\ 
\thetabh&\sim&{p(\thetab|\zbh,\xb,K,\gamma)}
\earr\right.
\eeq
where
\beq
p(\zb|\thetabh,\xb,K,\gamma)\propto p(\xb|\thetabh,\zb,K,\gamma)\; \pi(\zb|\gammab,K)
\eeq
and
\beq
p(\thetab|\zbh,\xb,K,\gamma)\propto p(\xb|\thetabh,\zb,K,\gammab)\;\pi(\thetab)
\eeq
where $\pi(\zb|\gamma,K)$ is given either by (\ref{Prior1}) or by  (\ref{Prior2}). 

The main difficulty in these relations is that the joint distribution $p(\zb,\thetab|\xb,K,\gamma)$ 
is not separable in its arguments. A framework which will give us the possibility to establish 
interesting relations between these approaches is the approximation of this joint distribution by a separable one which becomes variational techniques. 

\section{Variational Bayes}
To be able to compare the two approaches, we consider 
\beq
p(\zb,\thetab|\xb,K)=p(\xb|\zb,\thetab_1,K) \; \pi(\zb|\thetab_2,K) \; \pi(\thetab|K) / p(\xb|K)
\eeq
where $\thetab_1=\{\mub, \vb\}$, 
$\pi(\thetab_1)=\pi(\mub) \, \pi(\vb)=\sum_k \pi(\mu_k) \, \pi(v_k)$ with 
$\pi(\mu_k)=\Nc(\mu_k|\mu_0,v_0)$ and $\pi(v_k)=\Ic\Gc(v_k|\alpha_0,\beta_0)$ and where 
$\pi(\zb|\thetab_2,K)$ is given either by 
\beq \label{Prior1}
\pi(\zb|\alphab,K)=\prod_i P(Z_i=k|\alphab,K)=\prod_{k=1}^K \alpha_k^{\sum_i \delta(z_i-k)},
\eeq
where $\thetab_2=\alphab$, 
or by
\beq \label{Prior2}
\pi(\zb|\gamma,K)=\prod_i P(Z_i=k|\zb_{-i},\alphab,K)\propto \prod_i \exp\left\{\gamma \sum_{j\in\Vc(i)} \delta(z_i-z_j)\right\}
\eeq
where $\thetab_2=\gamma$ and where $\pi(\thetab|K)=\pi(\thetab_1|K)\; \pi(\thetab_2|K)$.

In these equations 
\beq
p(\xb|K)
=\sum_{\zb} \int p(\xb,\zb,\thetab|K) \d{\thetab}
=\sum_{\zb} \int p(\xb|\zb,\thetab_1,K) \; \pi(\zb|\thetab_2,K) \; \pi(\thetab|K) \d{\thetab}
\eeq
which is the evidence of the model $K$ and can be used to determine $K$. 

Let consider a free distribution $q(\zb,\thetab)$ and compare it to the joint posterior $p(\zb,\thetab|\xb,K)$ and the complete dat likelihood $p(\xb,\zb,\thetab|K)$ via the the two following quantities:
\bit
\item Free energy:
\beq
\barr{lcl}
\Fc\left(q(\zb,\thetab):p(\xb,\zb,\thetab|K)\right)
&=&\sum_{\zb} \int q(\zb,\thetab) \ln \frac{p(\xb,\zb,\thetab|K)}{q(\zb,\thetab)} \d{\thetab}  \\
&=&\sum_{\zb} \int q(\zb,\thetab) \ln \frac{p(\zb,\thetab|\xb,K) \; p(\xb|K)}{q(\zb,\thetab)} \d{\thetab} \\
&=&\sum_{\zb} \int q(\zb,\thetab) \ln \frac{p(\zb,\thetab|\xb,K)}{q(\zb,\thetab)} \d{\thetab} + \ln p(\xb|K) 
\earr
\eeq
\item Kullback-Leibler relative entropy 
between the free distribution $q(\zb,\thetab)$ and the joint posterior $p(\xb,\zb|\thetab,K)$:
\beq
\Kc\left(q(\zb,\thetab) : p(\xb,\zb|\thetab,K)\right)
=\sum_{\zb} \int q(\zb,\thetab) \ln \frac{q(\zb,\thetab)}{p(\zb,\thetab|\xb,K)} \d{\thetab} 
\eeq
\eit
Then, we may note that
\beq
\ln p(\xb|K)- \Fc\left(q(\zb,\thetab):p(\xb,\zb,\thetab|K)\right)
=\Kc\left(q(\zb,\thetab) : p(\xb,\zb|\thetab,K)\right)\ge 0
\eeq
so that the free energy $\Fc\left(q(\zb,\thetab):p(\xb,\zb,\thetab|K)\right)$ 
is a lower bound for 
$\ln p(\xb|K)$. 
This also shows that minimizing 
$\Kc\left(q(\zb,\thetab) : p(\xb,\zb|\thetab,K)\right)$ or maximizing $\Fc\left(q(\zb,\thetab)\right)$ result to the same optimal solution 
$q(\zb,\thetab) = p(\zb,\thetab|\xb,K)$ which is the joint posterior. 

These relations are valid for any 
$q(\zb,\thetab)$ and in particular for a separable $q(\zb,\thetab)=q_1(\zb)\; q_2(\thetab)$. This remark is the main idea behind the variational Bayes method which tries to approximate the joint non-separable 
distribution $p(\zb,\thetab|\xb,K)$ by a separable 
\(
q(\zb,\thetab|\xb,K)=q_1(\zb) \; q_2(\thetab)
\) 
where $q_1$ and $q_2$ have to be determined in such a way that either the  Kullback-Leibler criterion 
$\Kc(q:p)$ be minimized or the free energy $\Fc\left(q(\zb,\thetab)\right)$ be maximized. 
Noting that 
\[
\barr{lcl}
\Kc\left(q_1(\zb) q_2(\thetab):p(\zb,\thetab|\xb,K)\right)
&=&\sum_{\zb} q_1(\zb) \left(\int q_2(\thetab) \ln \frac{q_2(\thetab)}{p(\zb,\thetab|\xb,K)} \d{\thetab}\right) \\ 
&=&\sum_{\zb} q_1(\zb) \left(<p(\zb,\thetab|\xb,K)>_{q_2(\thetab)}-\Hc(q_2)\right)
\\ 
&=&\int q_2(\thetab) \left(\sum_{\zb} q_1(\zb) \ln \frac{q_1(\zb)}{p(\zb,\thetab|\xb,K)} \right) \d{\thetab}\\
&=&\int q_2(\thetab)  \left(<p(\zb,\thetab|\xb,K)>_{q_1(\zb)}-\Hc(q_1)\right)
\earr
\]
and the fact that $\Kc(q:p)$ is concave in $q_1$ for fixed $q_2$ and in $q_2$ for fixed $q_1$ 
its optimization can be done in an iterative way 
\[
\left\{\barr{lclcl}
\qh_1^{(t+1)}&=&\argmin{q_1}{\Kc(q_1 \; \qh_2^{(t)} : p)}&=&\argmax{q_1}{\Fc(q_1 \; \qh_2^{(t)} : p)}
\\
\qh_2^{(t+1)}&=&\argmin{q_2}{\Kc(\qh_1^{(t)} \; q_2 : p)}&=&\argmax{q_2}{\Fc(\qh_1^{(t)} \; q_2 : p)}
\earr\right.
\]
where $t$ notes the iteration number. It is then easy to show that, at each iteration $t$, the solutions is obtained by computing the derivatives of the corresponding functionals and equating them to zero, which leads to: 

\beq
\left\{\barr{lcl}
\qh_1^{(t+1)}(\zb)
&\propto& \expf{<\ln p(\xb,\zb,\Thetab|K)>_{q_2^{(t)}(\thetab)}}
\\
\qh_2^{(t+1)}(\thetab)
&\propto& 
\expf{<\ln p(\xb,\Zb,\thetab|K)>_{q_1^{(t)}(\zb)}}
\earr\right.
\eeq
where 
$<.>_q$ mean the expectation over $q$. 
For more details on this approach see \cite{Dempster77,Ghahramani97}. 

Noting that $p(\xb,\zb,\thetab|K)=p(\xb|\zb,\thetab,K)\, \pi(\zb|\thetab,K)\, \pi(\thetab|K)$, 
we see that the choice of the priors $\pi(\zb|\thetab,K)$ and $\pi(\thetab|K)$ as well as the choice of parametric 
family of $q_1(\zb)$ and $q_2(\thetab)$ is of great importance for the expressions of  $\qh_1^{(t+1)}(\zb)$ and $\qh_2^{(t+1)}(\thetab)$ 
and their final $\qh_1^{*}(\zb)$ and $\qh_2^{*}(\thetab)$. 
To obtain a computationally effective inference method, it is necessary to choose appropriately these distributions. For example, choosing conjugate priors for the hyperparameters $\pi(\thetab|K)$, we gain the advantage that the posterior $\pi(\thetab|\zb,\xb,K)$ or $\pi(\thetab|\xb,K)$ expressions will be in the same 
family than the associated priors. 

Between particular cases, we may mention the following:

\smallskip\noindent{\bf Optimal case:} 
\quad  $q_1(z)=p(\zb|\xb,K)$ \quad and \quad $q_2(\thetab)=p(\thetab|\xb,K)$ \\ 
This means that $p(\zb,\thetab|\xb,K)$ is approximated by the product 
$p(\zb|\xb,K) \, p(\thetab|\xb,K)$. The solution in this case is immediate:
\[
\left\{\barr{l} 
\qh_1^*(\zb)=p(\zb|\xb,K)={\sum_\zb p(\xb|\zb,\thetab,K) \, \pi(\zb|\thetab,K)} / {p(\xb|K)}\\ 
\qh_2^*(\thetab)=p(\thetab|\xb,K)={\int p(\xb|\zb,\thetab,K) \, \pi(\thetab|K) \d{\thetab}} / {p(\xb|K)}
\earr\right.
\]
However, computing any of these two terms needs integration (integration over $\thetab$ for the first and summation over $\zb$ for 
the second. 

\smallskip\noindent{\bf Degenerate case:}  
\quad  $q_1(\zb)=p(\zb|\thetabh^*,\xb,K)$ \quad and \quad $q_1(\thetab)=p(\thetab|\zbh^*,\xb,K)$. \\
where $\thetabh^*$ and $\zbh^*$ are two point estimators of $p(\thetab|\zb,\xb,K)$ and $p(\zb|\thetab,\xb,K)$. \\ 
This case is obtained through the following iterations:
\[
\left\{\barr{l} 
\qh_1^{(t)}(\zb)=\delta(\zb-\zbh^{(t)})\\ 
\qh_2^{(t)}(\thetab)=\delta(\thetab-\thetabh^{(t)})
\earr\right.
\lra
\left\{\barr{l} 
\qh_1^{(t+1)}(\zb)=p(\zb|\thetabh^{(t)},\xb,K)\\ 
\qh_2^{(t+1)}(\thetab)=p(\thetab|\zbh^{(t)},\xb,K)
\earr\right.
\]
which means that $p(\zb,\thetab|\xb,K)$ is approximated by the product 
$p(\zb|\thetab,\xb,K) \, p(\thetab|\zb,\xb,K)$. Both expressions are available up to their normalizing factors: 
\[
\left\{\barr{l} 
p(\zb|\thetab,\xb,K)\propto p(\xb|\zb,\thetab,K)\, \pi(\zb|\thetab,K)\\ 
p(\thetab|\zb,\xb,K)\propto p(\xb|\zb,\thetab,K)\, \pi(\thetab,K)
\earr\right.
\]
However, computing at each iteration $\thetabh^{(t)}$ and $\zbh^{(t)}$, which may be either the means or modes of these two distributions, may still need some effort. In particular, in the expression of $p(\zb|\thetab,\xb,K)$, depending on the prior $\pi(\zb|\thetab,K)$ the computational cost and difficulties are different. 
The separable case of (\ref{Prior1}) is much easier than the Markovian case of (\ref{Prior2}). 

\noindent{\bf Variational EM:}
\\ $q_1(\zb)=p(\zb|\thetabh^*,\xb,K)$ \quad and \quad $q_2(\thetab)=p(\thetab|\xb,K)$ \quad 
where \quad $\thetabh^*=\argmax{\thetab}{p(\thetab|\xb,K)}$. \\ 
This case is obtained through the following iterations:
\[
\qh_2^{(t)}(\thetab)=\delta(\thetab-\thetabh^{(t)})
\lra
\left\{\barr{ll} 
\qh_1^{(t+1)}(\zb)=p(\zb|\thetabh^{(t)},\xb,K)\\ 
\thetabh^{(t+1)}=\argmax{\thetab}{Q(\thetab,\thetabh^{(t)})} \mbox{~~with} 
& \mbox{(M step)}\\ 
Q(\thetab,\thetabh^{(t)})=<\ln p(\xb,\Zb,\thetab|K)>_{p(\zb|\thetabh^{(t)},\xb,K)}
& \mbox{(E step)}
\earr\right.
\]

The next step in approximations is to choose  
$q_1(\zb)=\prod_i q_{1i}(z_i)$ or $q_2(\thetab)=\prod_k q_{2k}(\theta_k)$ or both. 
The first case is only necessary for Markovian models of the labels. 

\smallskip\noindent{\bf Mean Field + EM:}
\\ $q_1(\zb)=\prod_i p(z_i|\zb_{-i},\thetabh^*,\xb,K)$ ~and~ $q_2(\thetab)=p(\thetab|\xb,K)$  
~where~ $\thetabh^*=\argmax{\thetab}{p(\thetab|\xb,K)}$. \\ 

This case is obtained through the following iterations:
\[
\left\{\barr{l} 
q_1(\zb)=\prod_i q_{1j}(z_i) \\ 
\qh_2^{(t)}(\thetab)=\delta(\thetab-\thetabh^{(t)}) 
\earr\right.
\lra
\left\{\barr{ll} 
\qh_{1j}^{(t+1)}(z_i|\zb_{-i})\propto p(z_i|\zb_{-i},\thetabh^{(t)},\xb,K) \\
\thetabh^{(t+1)}=\argmax{\thetab}{Q(\thetab,\thetabh^{(t)})} \mbox{~~with}
& \mbox{(M step)}\\ 
Q(\thetab,\thetabh^{(t)})=<\ln p(\xb,\Zb,\thetab|K)>_{p(\zb|\thetabh^{(t)},\xb,K)}
& \mbox{(E step)}
\earr\right.
\]
where 
\[
p(z_i|\zb_{-i},\thetabh^{(t)},\xb,K)\propto 
p(\xb|\zb,\thetabh^{(t)},K) \, p(z_i|\zb_{-i},\thetabh^{(t)})
\]

\smallskip\noindent{\bf Mean Field + separable EM:}
\\ $q_1(\zb)=p(\zb|\thetabh^*,\xb,K)$ ~and~ $q_2(\thetab)=\prod_k q_{2k}(\thetab_k|\xb,K)$  
~where~ $\thetabh_k^*=\argmax{\theta_k}{p(\thetab_k|\xb,K)}$
where 
\[
\barr{ll} 
q_2(\thetab)=\prod_k q_{2k}(\alpha_k) \, q_{2k}(\mu_k) \, q_{2k}(v_k) &\mbox{for Dirichlet} \\ 
q_2(\thetab)=q_{2}(\gamma) \, \prod_k q_{2k}(\mu_k) \, q_{2k}(v_k) &\mbox{for Potts}
\earr
\]

\smallskip\noindent{\bf Totally separable (Mean Field):}
\\ $q_1(\zb)=\prod_i p(z_i|\zb_{-i},\thetabh^*,\xb,K)$ ~and~ 
$q_2(\thetab)=\prod_k  q_{2k}(\alpha_k) q_{2k}(\mu_k)  q_{2k}(v_k)$. \\ 

\section{Applications in data classification and in image segmentation}
The mixture of Gaussians model are natural models for data classification. When the data are the scalar grey level $x_i$ or the color components $\xb_i$ of a pixel in an image, their classification result to the segmentation of that image. In the following, we note by $\rb_i$ the coordinate position of the pixel i, by $x_i=x(\rb_i)$ its grey level, by $z_i=z(\rb_i)$ its classification label and by $\Rc_{kl}=\{r_i: z(\rb_i)=k\}$ all the disjoint compact regions having the same label.  
We assume that $\Rc_{kl}\cap\Rc_{kl'}=\phi$, $\Rc_{kl}\cap\Rc_{k'l}=\phi$ and  $\cup_k \Rc_{kl}\cap\Rc_{kl'}=\Rc$ which cover the whole image. What is more specifique in image segmentation compared to other data classification is the fact that there is a spatial organization of the data. A MoG model with Dirichlet prior does not account for this spatial organization, but the the same MoG with a Markovian Pottz prior accounts for that. There are also many other possibilities of modeling this spatial organization, but the Potts Markov model is the probably the simplest one. To illustrate this, let consider the image of the Figure 1-a and its original labels in b). The histogram of the image pixels c) is shown in c) and the histogram of the labels in d). The results of segmentation using MoG with Dirichlet prior are shown in e) supervised and in f) unsupervised. 
The results of segmentation using MoG with Potts prior are shown in g) supervised and in h) unsupervised. We observe that, in both cases of supervised and unsupervised, the results with Potts prior are better that those with Dirichlet priors.  

\bfig
\btabu{@{}c@{.}c|c@{.}c@{}}
\includegraphics[width=35mm,height=35mm]{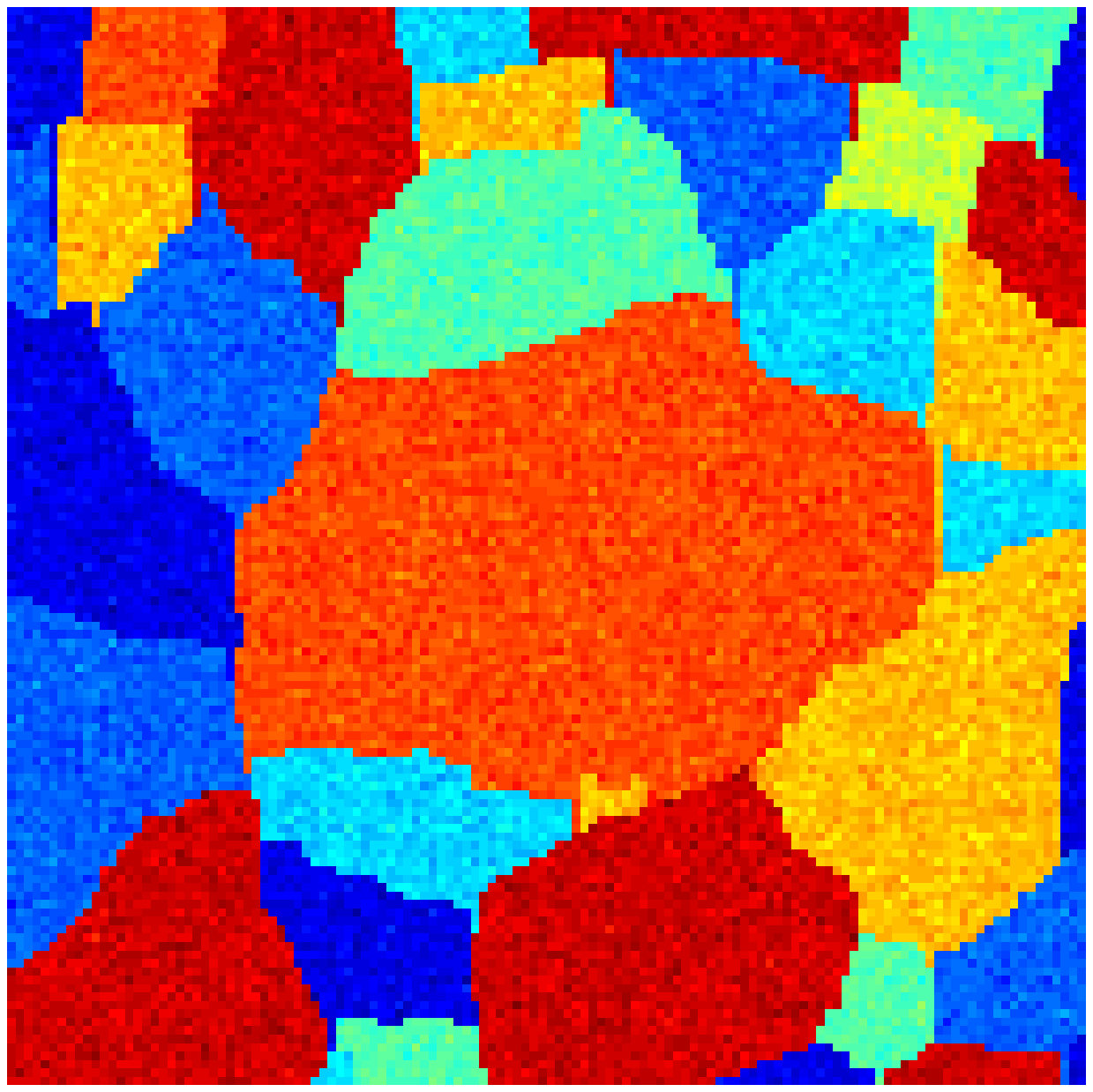}&
\includegraphics[width=35mm,height=35mm]{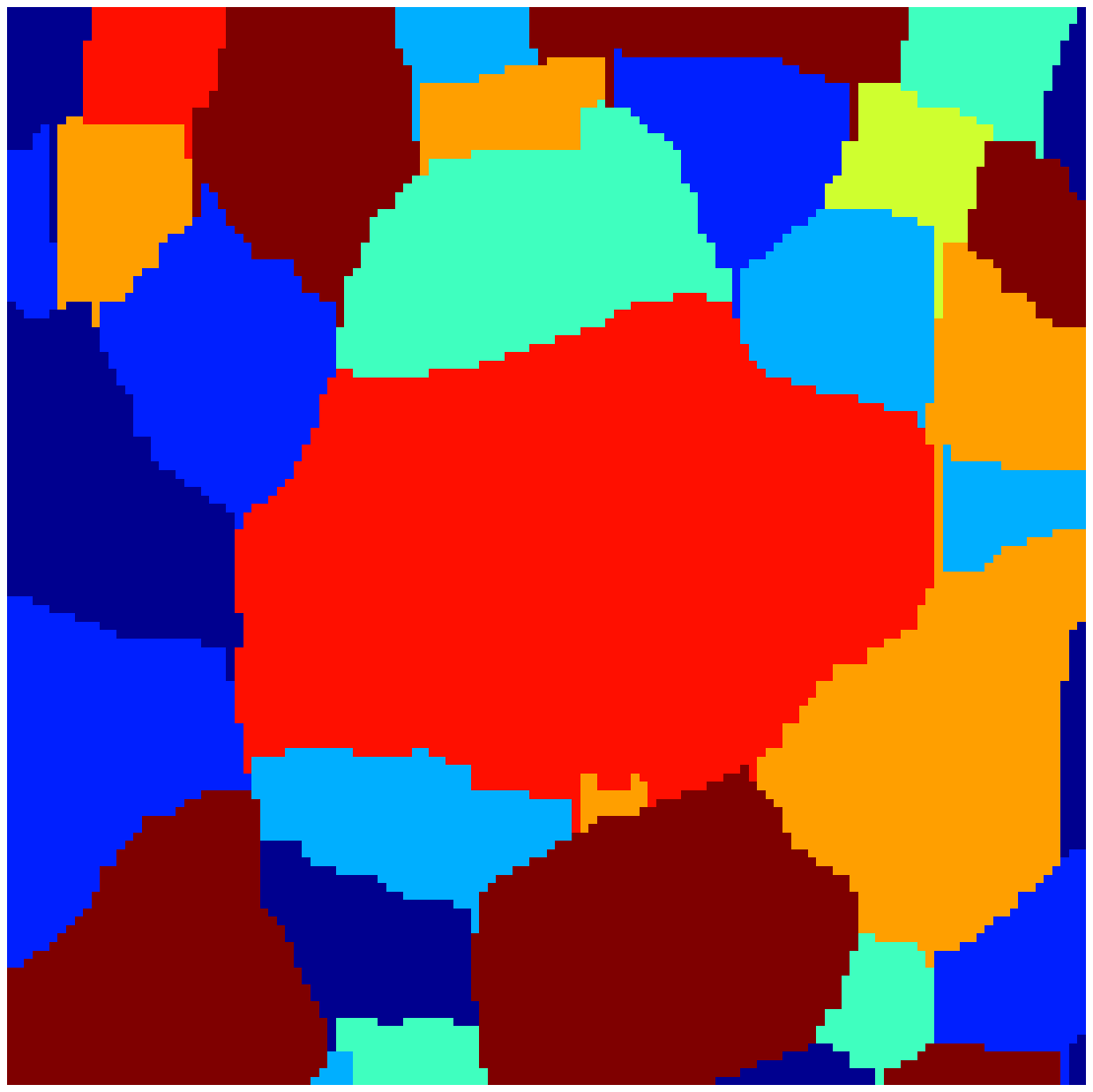}&
\includegraphics[width=35mm,height=35mm]{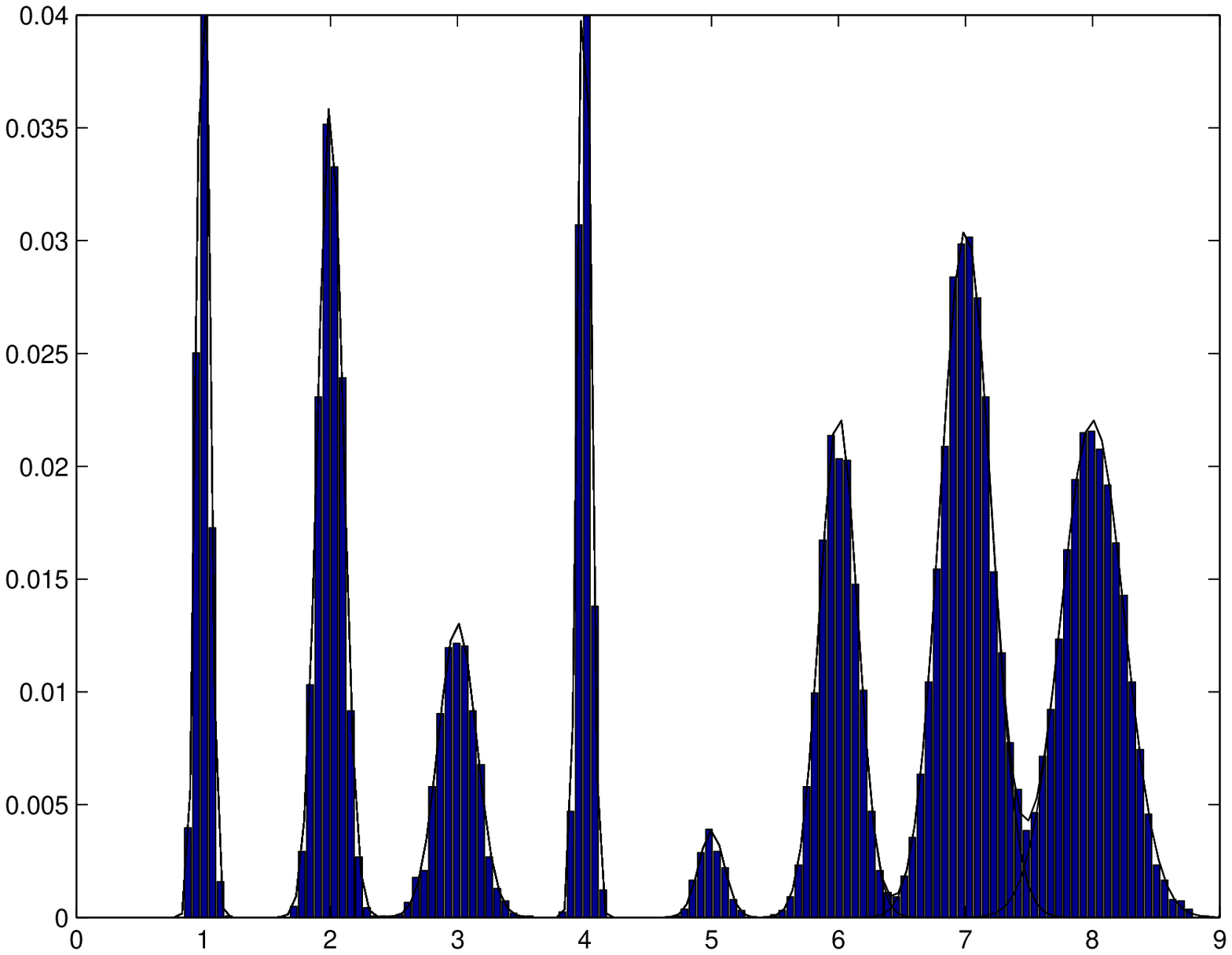}& 
\includegraphics[width=35mm,height=35mm]{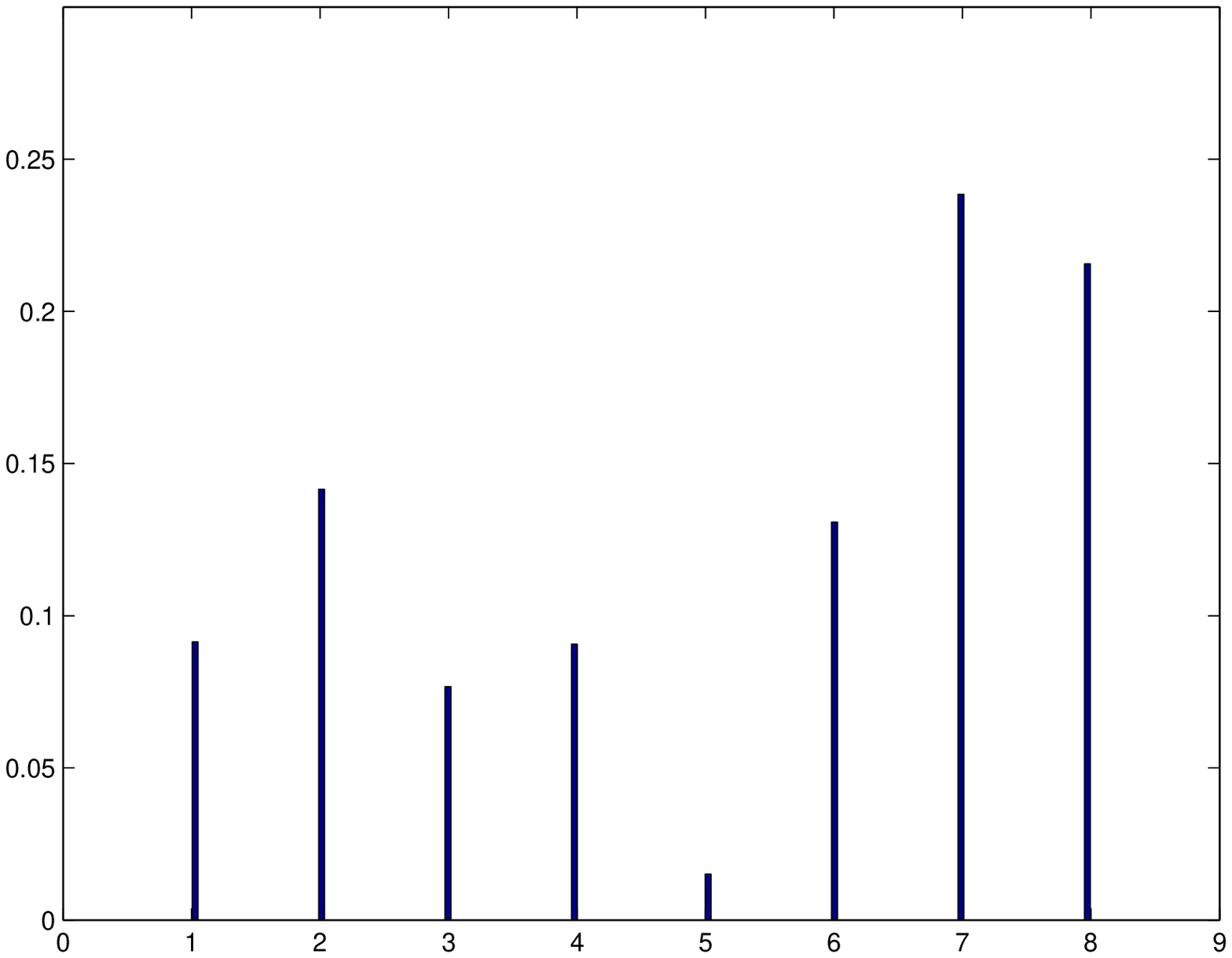}
\\ 
a) Original & b) labels & c) histogram  & d) histogram \\
            &           & of image      & of labels  
\\
\includegraphics[width=35mm,height=35mm]{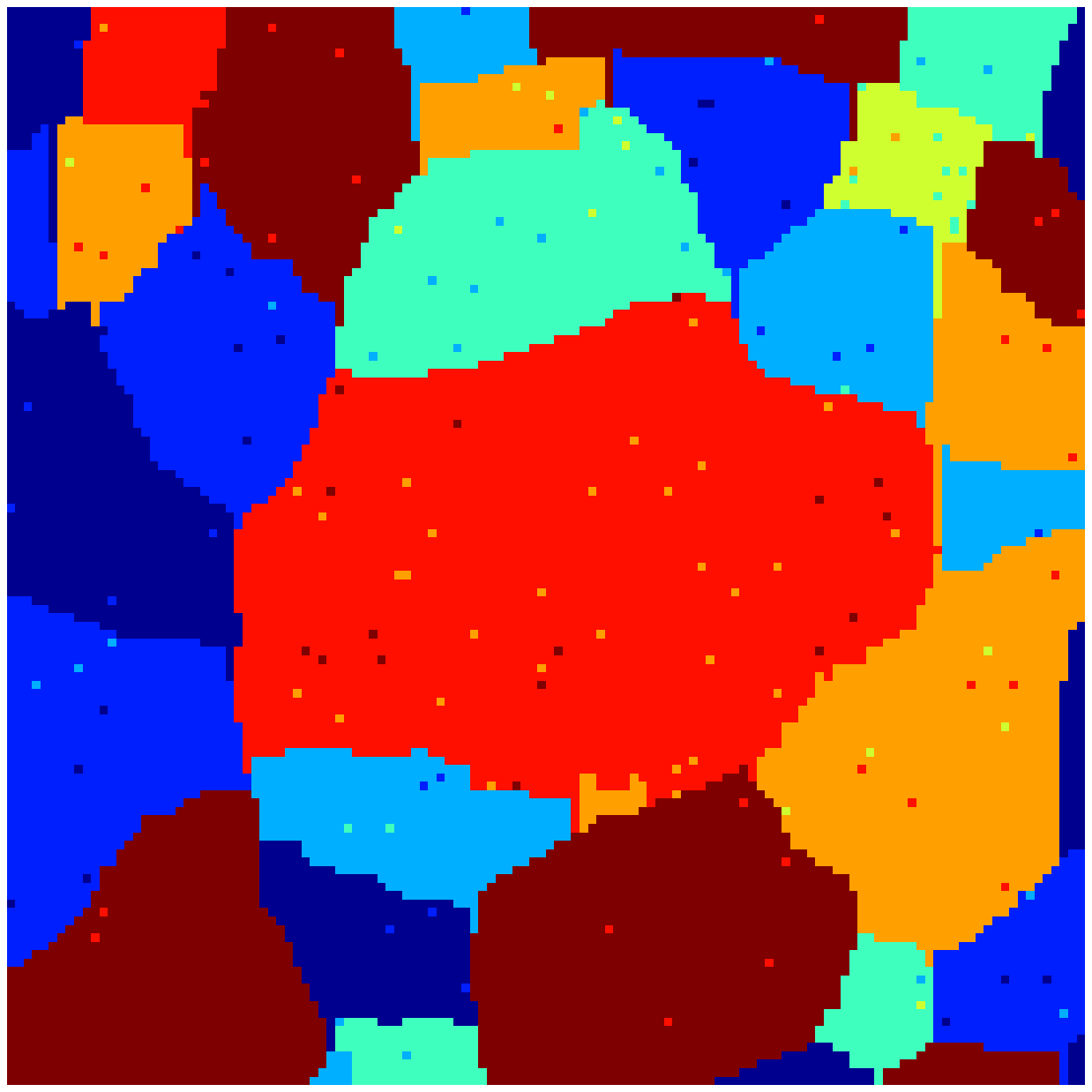}&
\includegraphics[width=35mm,height=35mm]{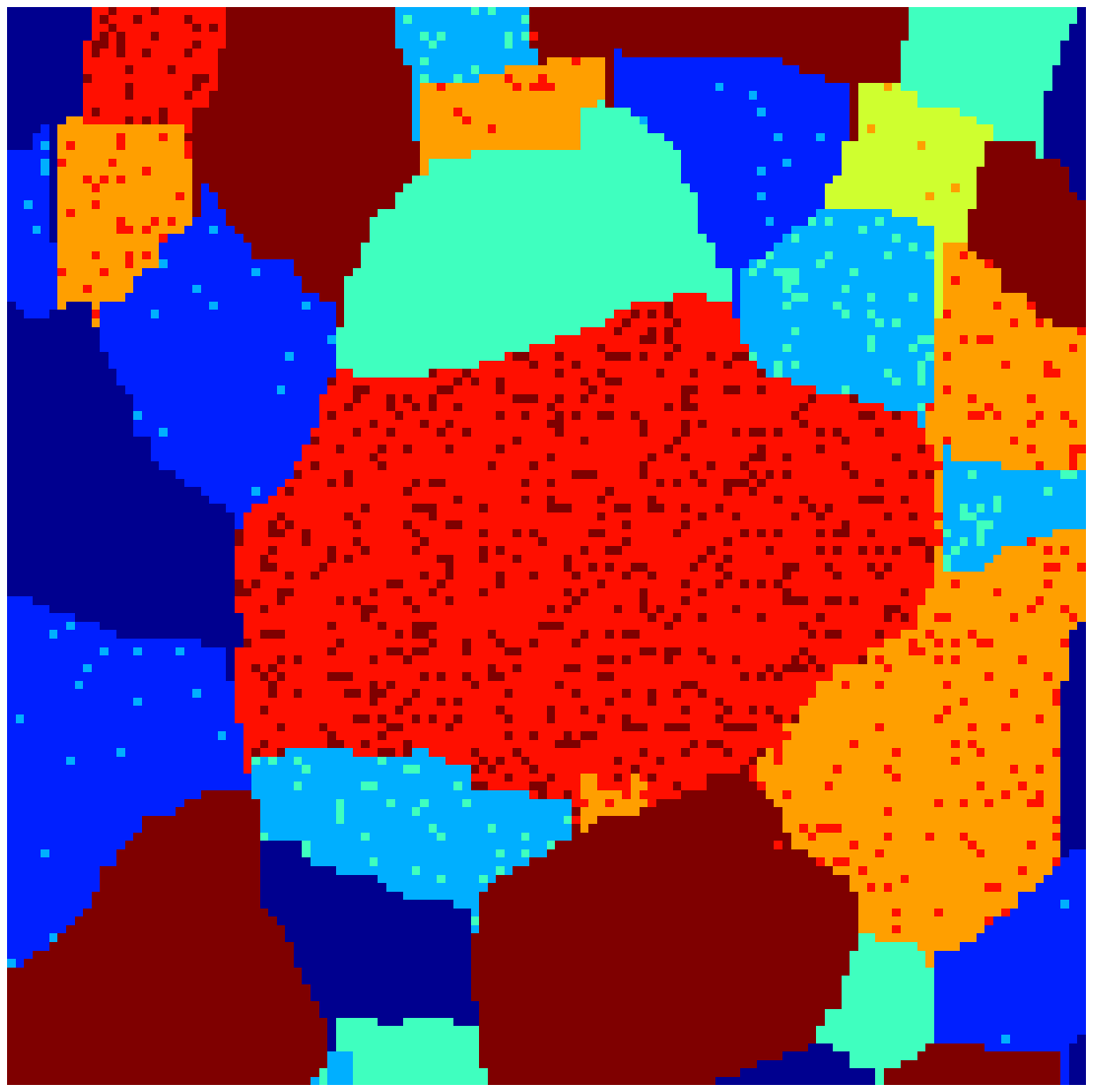}&
\includegraphics[width=35mm,height=35mm]{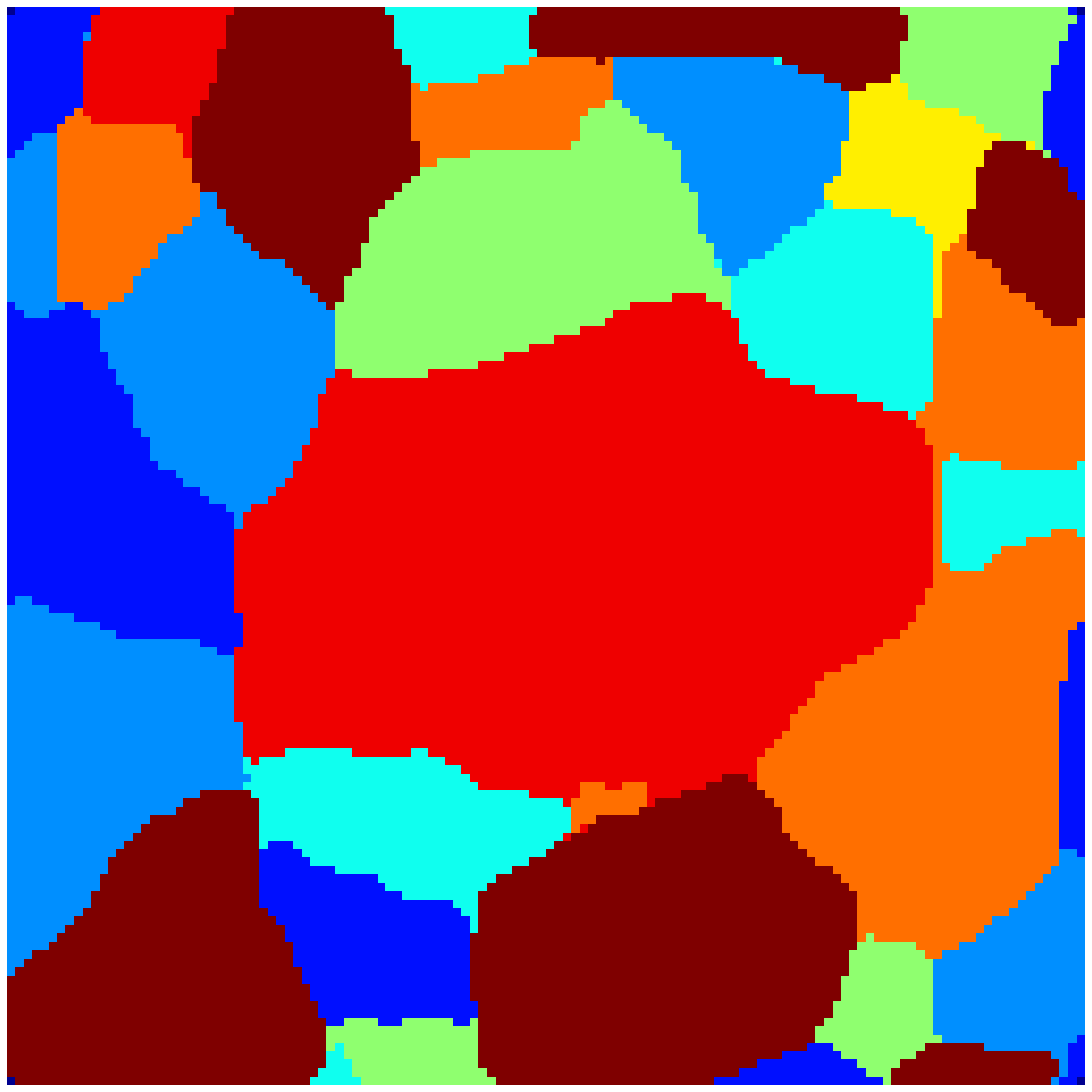}&
\includegraphics[width=35mm,height=35mm]{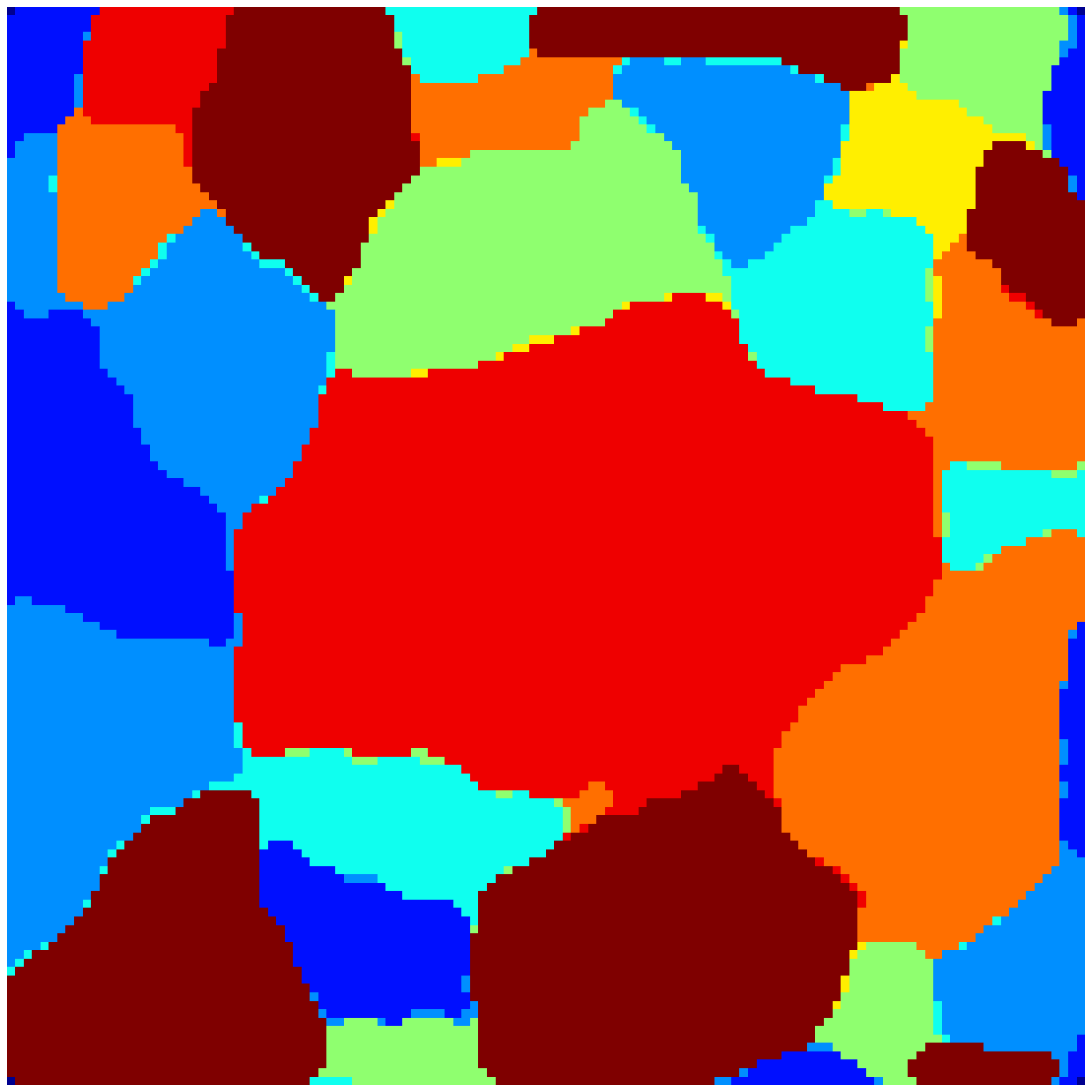} 
\\ 
e) supervised & f) unsupervised  & g) supervised & h) unsupervised \\
   Dirichlet  &    Dirichlet     &    Potts      &    Potts    
\\ 
\includegraphics[width=35mm,height=35mm]{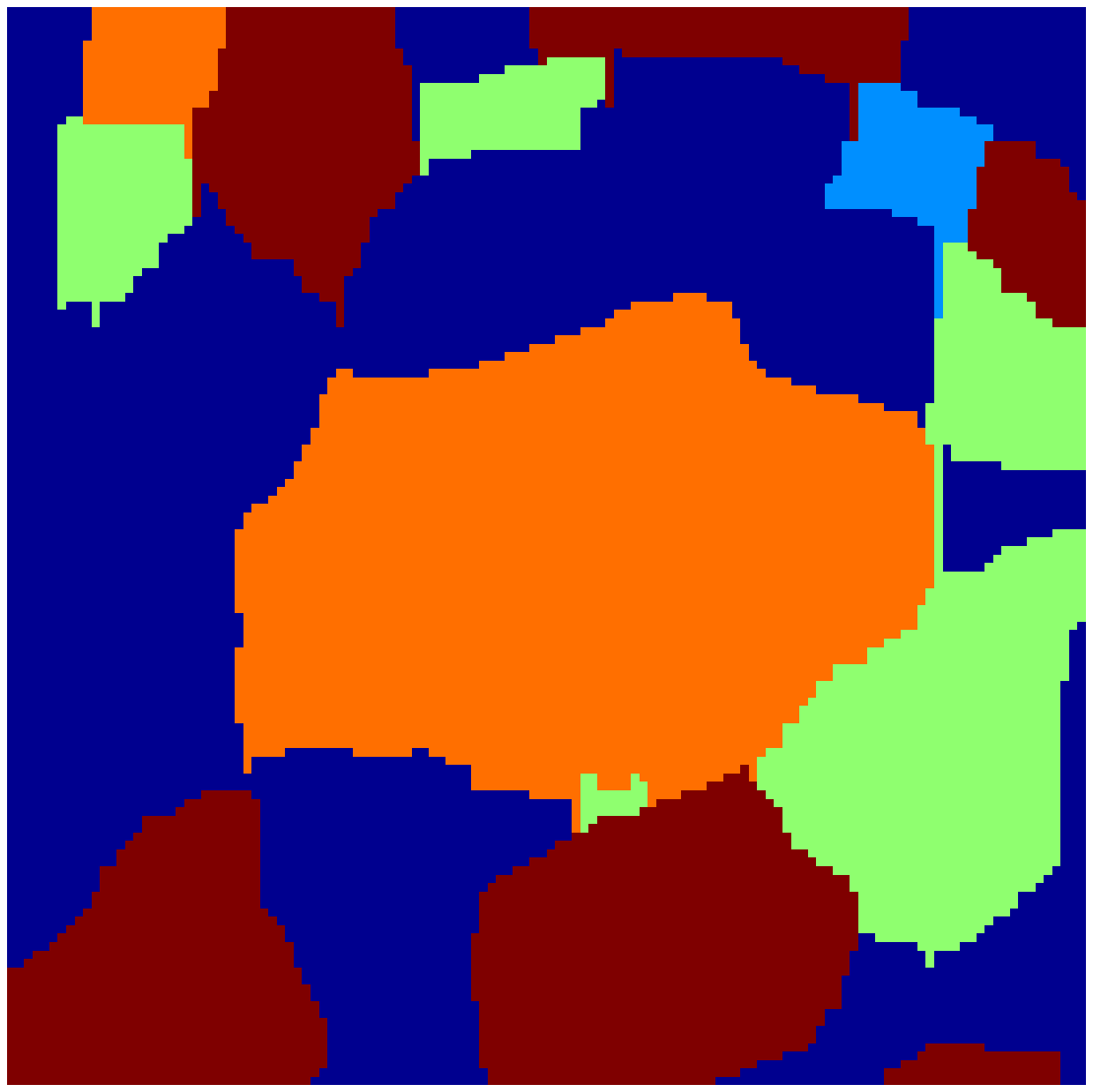}&
\includegraphics[width=35mm,height=35mm]{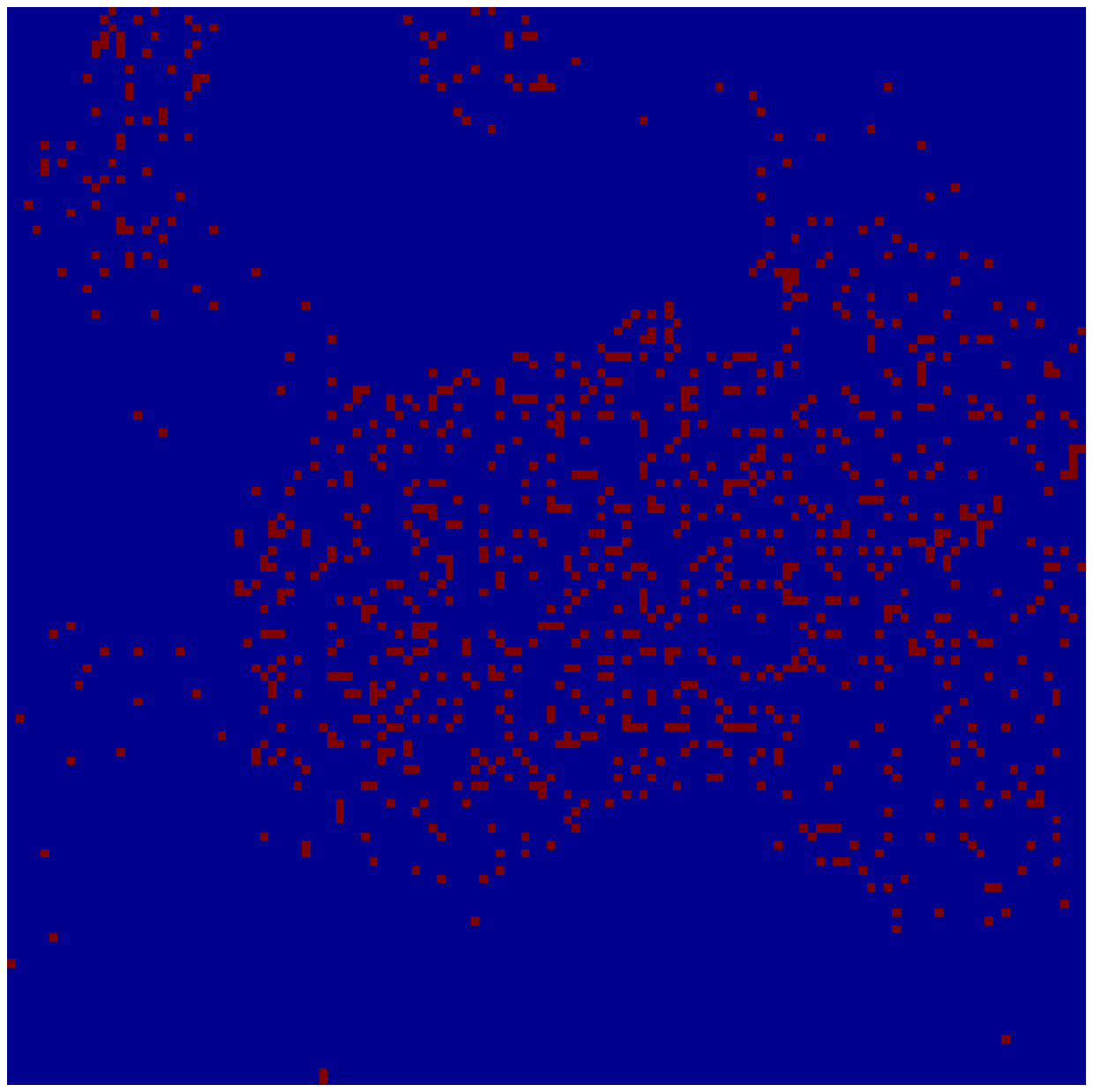}& 
\includegraphics[width=35mm,height=35mm]{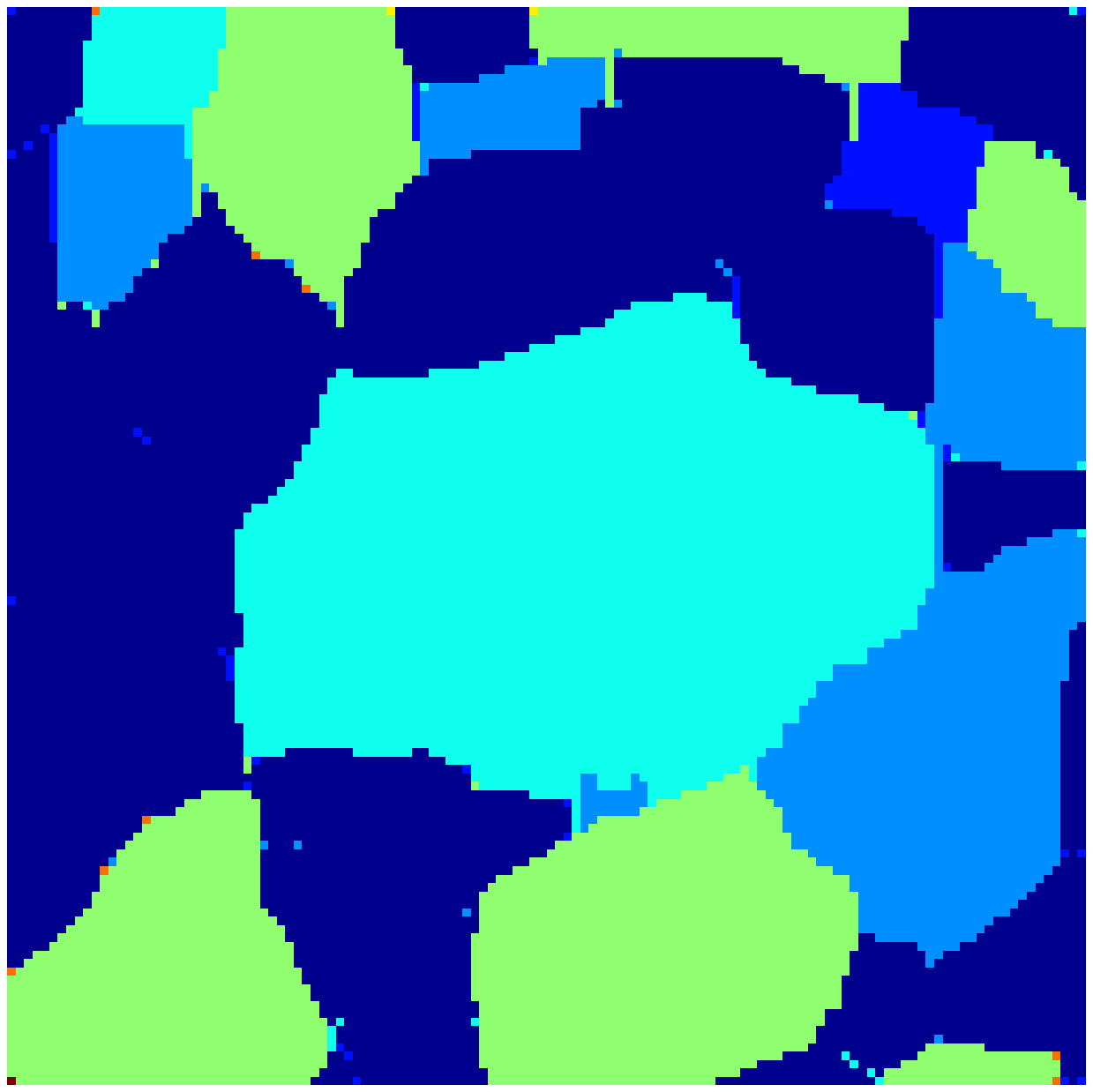}& 
\includegraphics[width=35mm,height=35mm]{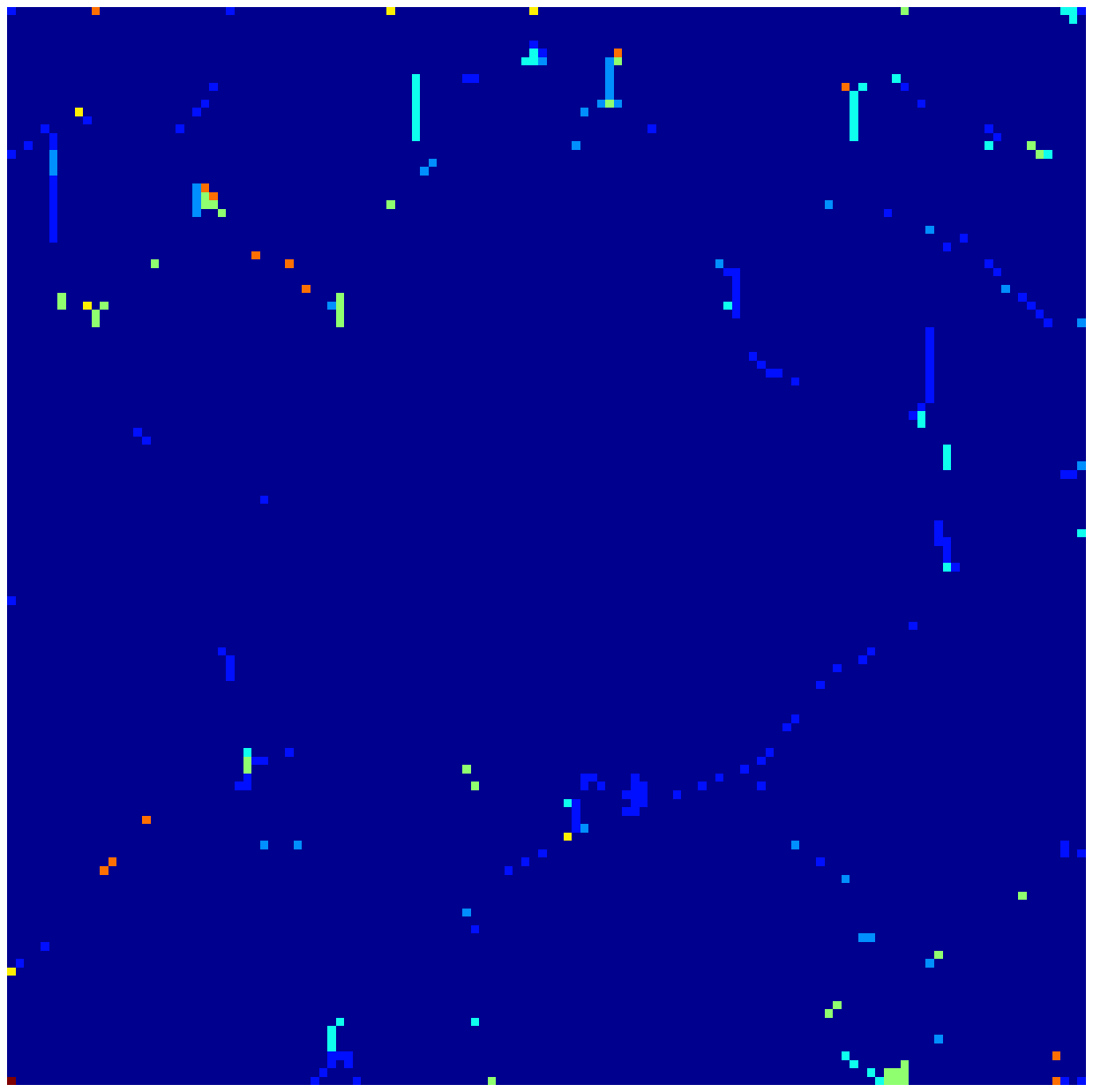}
\\ 
|e-b| & |f-b| & |g-b| & |h-b| \\ 
7.36\% & 20.0\% & 1.83\% & 4.64\%
\etabu
\caption{Classification and segmentation: 
a) an image, b) its original classification labels, c) histogram of the pixels, d) histogram of the labels, 
e) and f) Segmentation using MoG with Dirichlet (supervised and unsupervised; 
g) and h) Segmentation using MoG with Pottz (supervised and unsupervised);
i,j,k,l are the classification errors (differences between original classification in b) and obtained classification in e), f), g) and h).}
\label{Fig1}
\efig

\section{Conclusion}
The mixture of Gaussians model are used extensively for data classification and in image segmentation. When there is no prior knowledge of any spatial organization of the data, the Dirichlet prior can be used. However, in image segmentation, this model often does not give satisfactory results, because the spatial organization of the pixels is ignored. Using the Potts prior gives better results because this prior accounts for the spatial organization of the pixels.

\def\bibdir{/home/djafari/Tex/Inputs/bib/commun/}
\def\bibdirb{/home/djafari/Tex/Inputs/bib/amd/}
{\small 
\def\sca#1{{\sc #1}}
\bibliographystyle{ieeetr}
\bibliography{bibenabr,revuedef,revueabr,baseAJ,baseKZ,gpipubli,\bibdirb amd_art,\bibdirb amd_ca}
}
\end{document}